\documentclass[graybox]{svmult}
\usepackage{natbib}
\usepackage{type1cm}        % activate if the above 3 fonts are
\usepackage{makeidx}         % allows index generation
\usepackage{graphicx}        % standard LaTeX graphics tool
\usepackage{multicol}        % used for the two-column index
\usepackage[bottom]{footmisc}% places footnotes at page bottom

\usepackage{newtxtext}       % 
\usepackage{newtxmath}       % selects Times Roman as basic font

\makeindex             % used for the subject index
\def\apjl{ApJ}
\def\apj{Astrophys. J.}

\def\apjs{ApJS}

\def\mnras{MNRAS}

\def\aap{Astron. and Astrophys.}

\def\nat{Nature}

\def\aj{AJ}
\def\apss{Astrophysics and Space Science}

\newcommand{\be} {\begin{equation}}

\newcommand{\fermi}{{\emph{Fermi}}}

\newcommand{\bc}{\begin{center}}
\newcommand{\ec}{\end{center}}
\def\ltsima{$\; \buildrel < \over \sim \;$}
\def\lsim{\lower.5ex\hbox{\ltsima}}
\def\loe{\lower.5ex\hbox{\ltsima}}
\def\gtsima{$\; \buildrel > \over \sim \;$}
\def\gsim{\lower.5ex\hbox{\gtsima}}
\def\goe{\lower.5ex\hbox{\gtsima}}

\def\hess {HESS\,J0632$+$057}

\def\lsi {LS I\,$+$61 303}

\begin{document}

\title*{The high-energy emission of millisecond pulsars}

\author{Diego F. Torres and Jian Li }

\institute{Diego F. Torres \at ICREA \& Institute of Space Sciences (ICE, CSIC), \& Institut d'Estudis Espacials de Catalunya, c.can Magrans, s/n, 08193, Bellaterra, Barcelona, \email{dtorres@ice.csic.es}
\and Jian Li \at Deutsches Elektronen Synchrotron DESY, D-15738 Zeuthen, Germany \email{jian.li@desy.de}}


\maketitle

\abstract{This chapter provides a phenomenological appraisal of the high-energy emission of millisecond pulsars.
We comment on some of their properties as a population, as well as consider the especial cases of transitional pulsars, other redbacks, and 
black widow systems.
}

%%%%%%%%%%%%%%%%%%%%%%%%%%%%%%%%%%%%%%%
\section{Millisecond pulsars in the {\it Fermi}-LAT era}
%%%%%%%%%%%%%%%%%%%%%%%%%%%%%%%%%%%%%%%
\label{sec:1}

\begin{figure}[tb]
\includegraphics[scale=.6]{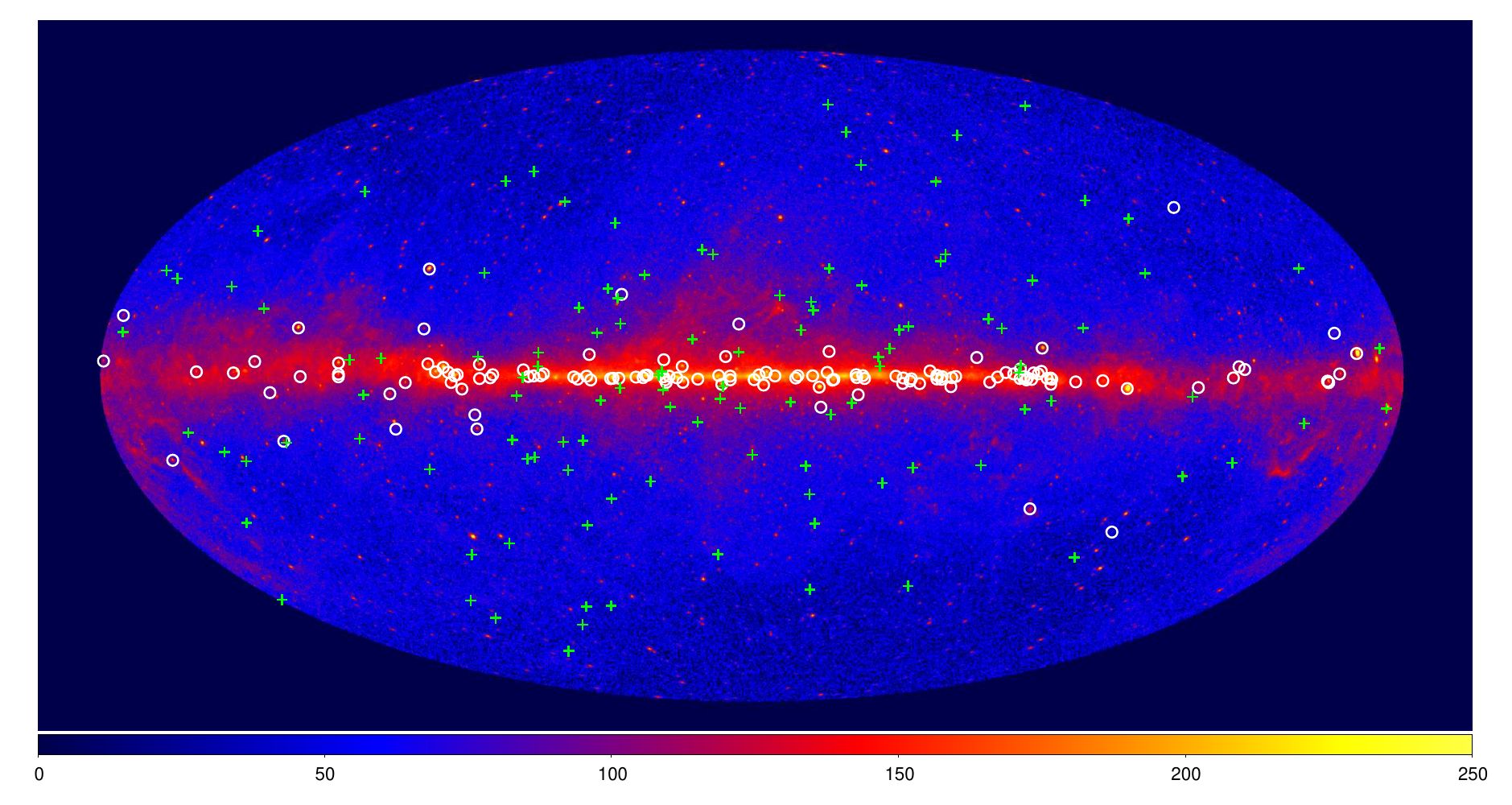}
\caption{Eight years of \fermi-LAT observations: all-sky map above 1 GeV.
Gamma-ray millisecond pulsars are shown with green crosses while other gamma-ray pulsars are shown with white circles.
The map in the background is adapted from \cite{smith2017}.
}
\label{map}       % Give a unique label
\end{figure}

The \fermi\ gamma-ray telescope was launched on June 11, 2008.
The main instrument onboard \fermi\ is the Large Area Telescope (LAT).
The LAT is an imaging high-energy gamma-ray telescope working in the energy range from 20 MeV to above 300 GeV.
The LAT has a large file of view (20\% sky coverage at any time), as well as a significant effective area (on the scale of 10$^{4}$ cm$^{2}$).
LAT is mostly operating in a survey mode, scanning the whole sky every three hours.
Because of its large effective area and field of view, maximized with a survey strategy, \fermi-LAT has been the main working horse in gamma-ray pulsar research since its launch.
During the three decades after the discovery of what was meant to be the first gamma-ray pulsar (i.e., the Crab pulsar, PSR B0531+21; \cite{Browning1971}), only seven pulsars were known to emit in gamma-rays.
Within only six months since its launch, Fermi-LAT has increased this number to 46 \citep{1PC}, and later to 117 in three years \citep{2PC}.
Until December 27, 2019, the number of \fermi-LAT detected gamma-ray pulsars reached 250\footnote{See the updated list at \url{https://confluence.slac.stanford.edu/display/GLAMCOG/Public+List+of+LAT-Detected+Gamma-Ray+Pulsars}}.
This population is composed of 115 millisecond pulsars (MSPs, $P<$30 ms) and 135 young, non-recycled pulsars  ($P>$30 ms).
Their positions are shown in Figure \ref{map}, on the background of the eight-years \fermi-LAT all sky map above 1 GeV (see e.g., \citep{smith2017}).
Gamma-ray MSPs (in short, referred to as gMSPs in what follows) are scattered at larger Galactic latitude in comparison to gamma-ray young pulsars, which are mainly distributed in the Galactic plane.
This is consistent with most of MSP being of recycled nature (neutron-star formation in an evolving binary system and spin-up due to accretion from the binary companion), leading to a larger age of their population.

This chapter provides a phenomenological appraisal of the high-energy emission of millisecond pulsars.
We comment on some of their properties as a population, as well as consider the especial cases of transitional pulsars, other redbacks, and 
black widow systems.
An interesting and complementary overview can be found in the work of \cite{smith2017}.
Other chapters in this same volume will deal with theoretical aspects of the high-energy emission from millisecond pulsars beyond what we mention here, see also
\cite{Venter2018}.

Besides the obviously faster spin period ($\dot{P}$) and much smaller value of period derivative ($\dot{P}$), many different characteristics have been revealed when comparing the population of gMSPs with that of the gamma-ray young pulsars observed by \fermi-LAT, as reported in the gamma-ray pulsar catalogs
\citep{1PC,2PC}.
Among these differences, we can emphasize that 
gMSPs have lower gamma-ray luminosities compared to gamma-ray young pulsars, which explains why gMSPs are often detected at nearer distances.
Gamma-ray pulsars are multi-wavelength objects and many  of them are bright also in X-rays.
The observed gamma-ray to X-ray flux ratio is lower for gMSPs, and it is less scattered than that observed for the gamma-ray young pulsars.
The spectra of gamma-ray pulsars are usually modelled by a power-law with exponentially cutoff.
The power-law index $\Gamma$ shows a mild correlation with the spin down power $\dot{E}$ for gMSPs and gamma-ray young, radio-quiet pulsars, with a Pearson correlation factor of 0.58 and 0.68, respectively.
The cutoff energy $E_{cut}$ correlates with the magnetic field strength at light cylinder, with a Pearson correlation factor of 0.52 for gMSPs and 0.64 for gamma-ray young, radio-quiet pulsars.
Again, some dispersion is found also here.
The Vela \citep{vela2018} and Crab \citep{Crab2016} pulsar, both of which are gamma-ray young pulsars, have also been detected by Imaging Atmospheric Cherenkov Telescopes at very high energies (VHE) above 100 GeV.
No gMSP has been detected in the VHE range so far.

\begin{table}[htp]
\caption{Gamma-ray millisecond pulsar varieties as of December 27, 2019.}
\begin{center}
\begin{tabular}{lll}
\hline\hline
\\
  Category    & $\#$     & Fraction \\
   \\
\hline\hline
\\
Known gamma-ray MSP    &  115               &       \\
\\
\hline
\\
\ldots with radio pulsation   &   112    &97.4\%                   \\
\ldots without radio pulsation  &  3        &2.6\%              \\
\\
\hline
\\
\ldots Isolated    &  24       & 20.9\%             \\
\ldots In binaries  &  91          & 79.1\%          \\
\\
\hline
\\
Redbacks &  10 &  \\
Black widows  & 21 &  \\
\\
\hline
\end{tabular}
\end{center}
\label{sum}
\end{table}%

\begin{figure}[tb]
\centering
\includegraphics[scale=.25]{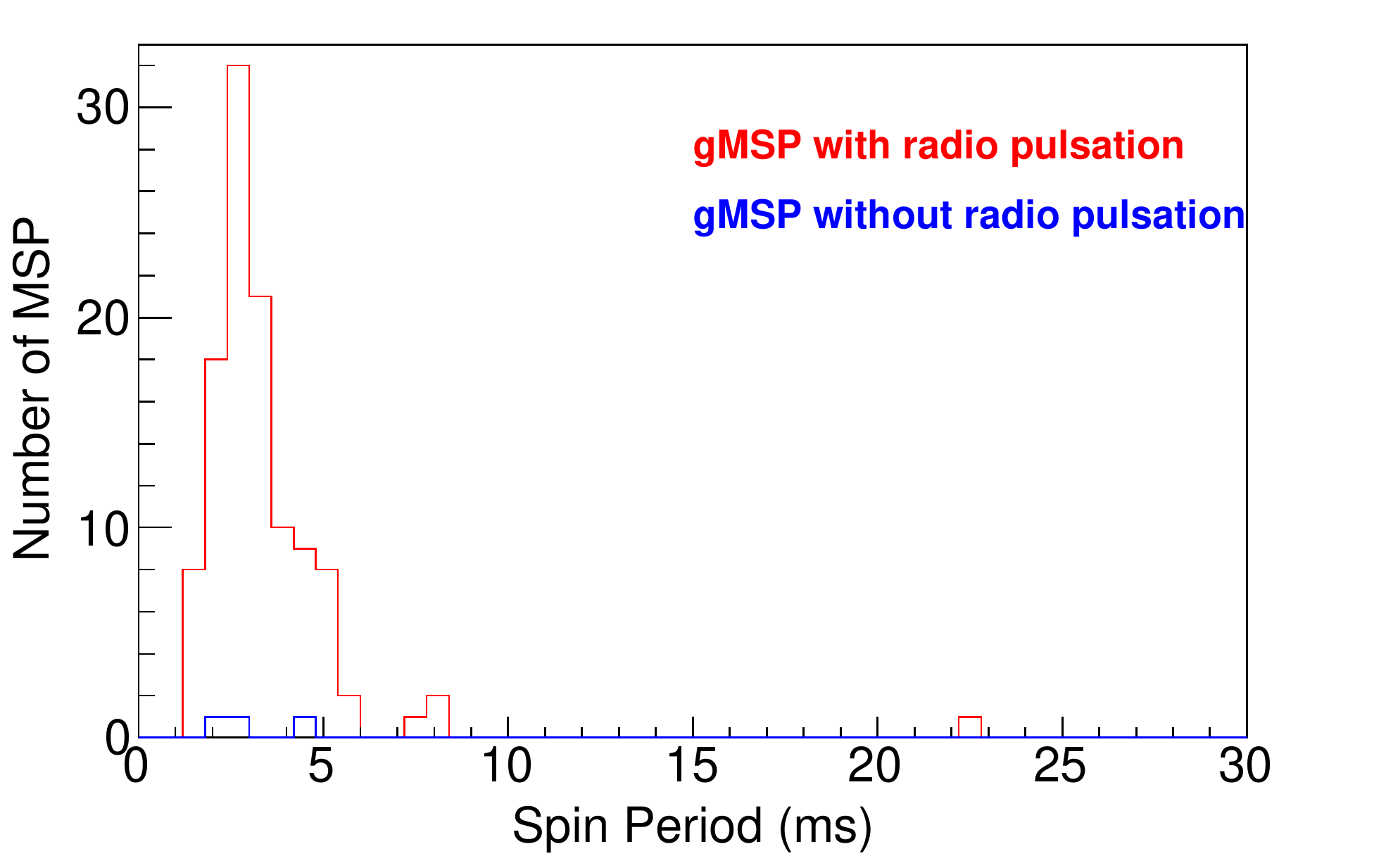}
\includegraphics[scale=.25]{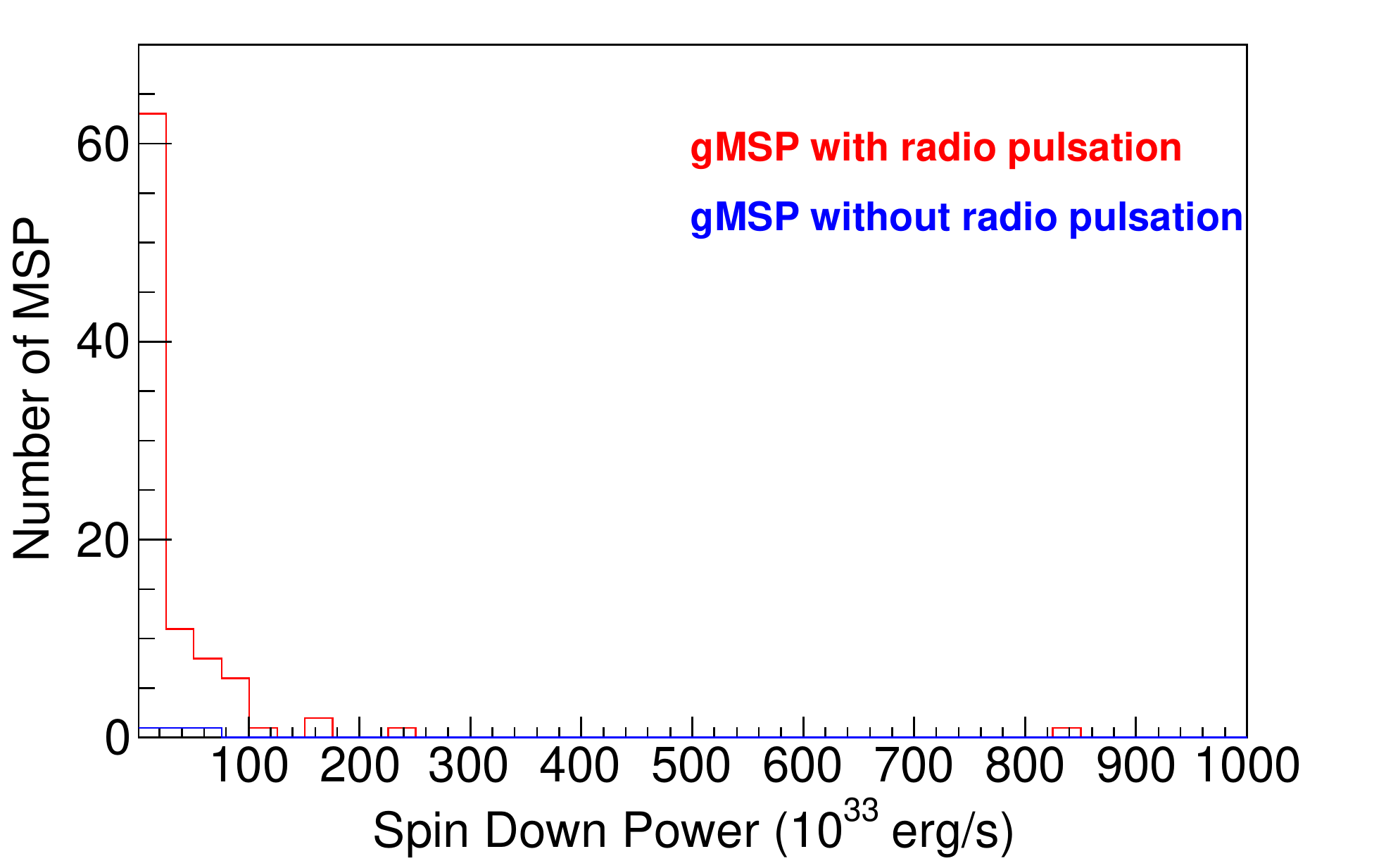}\\
\includegraphics[scale=.25]{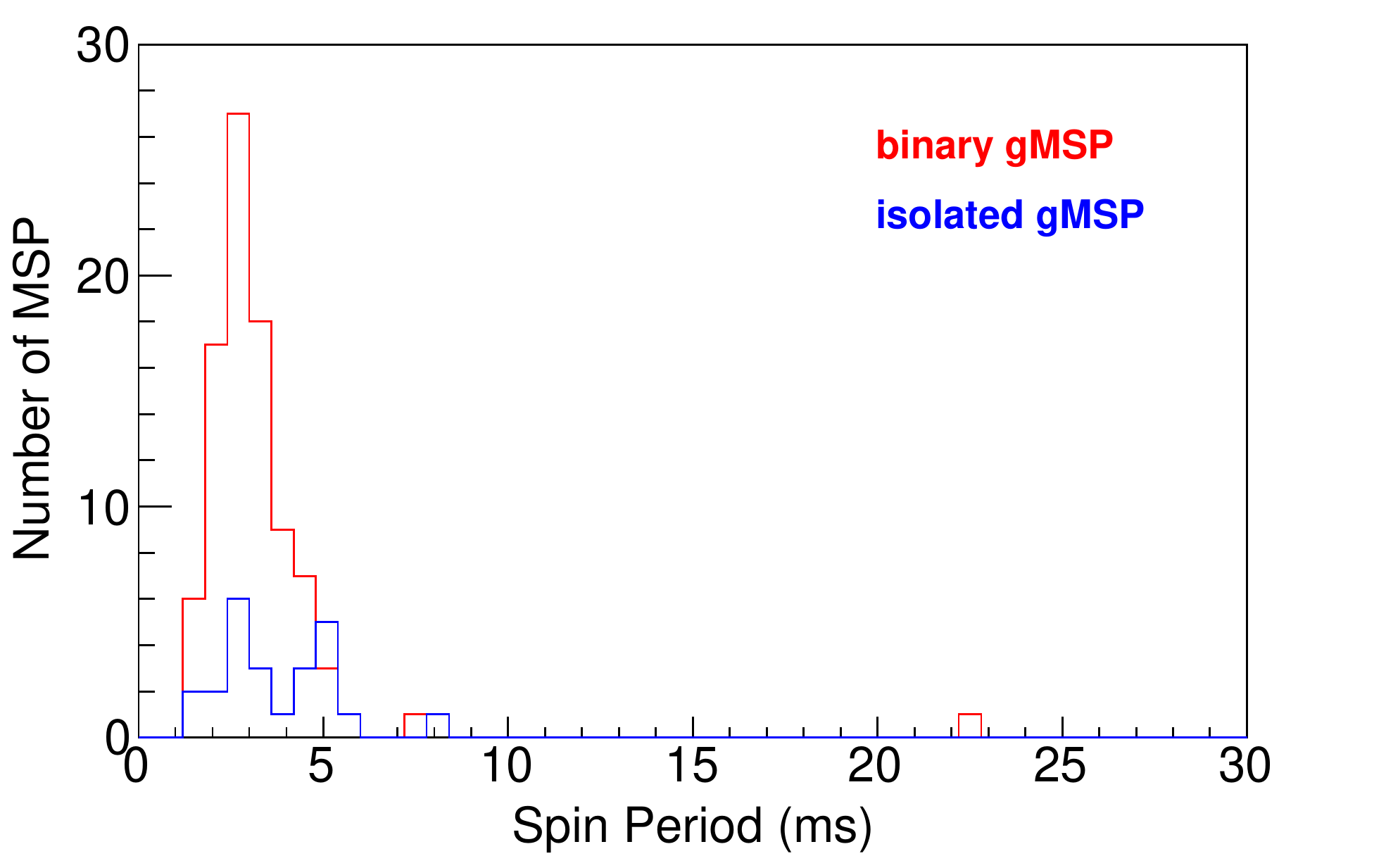}
\includegraphics[scale=.25]{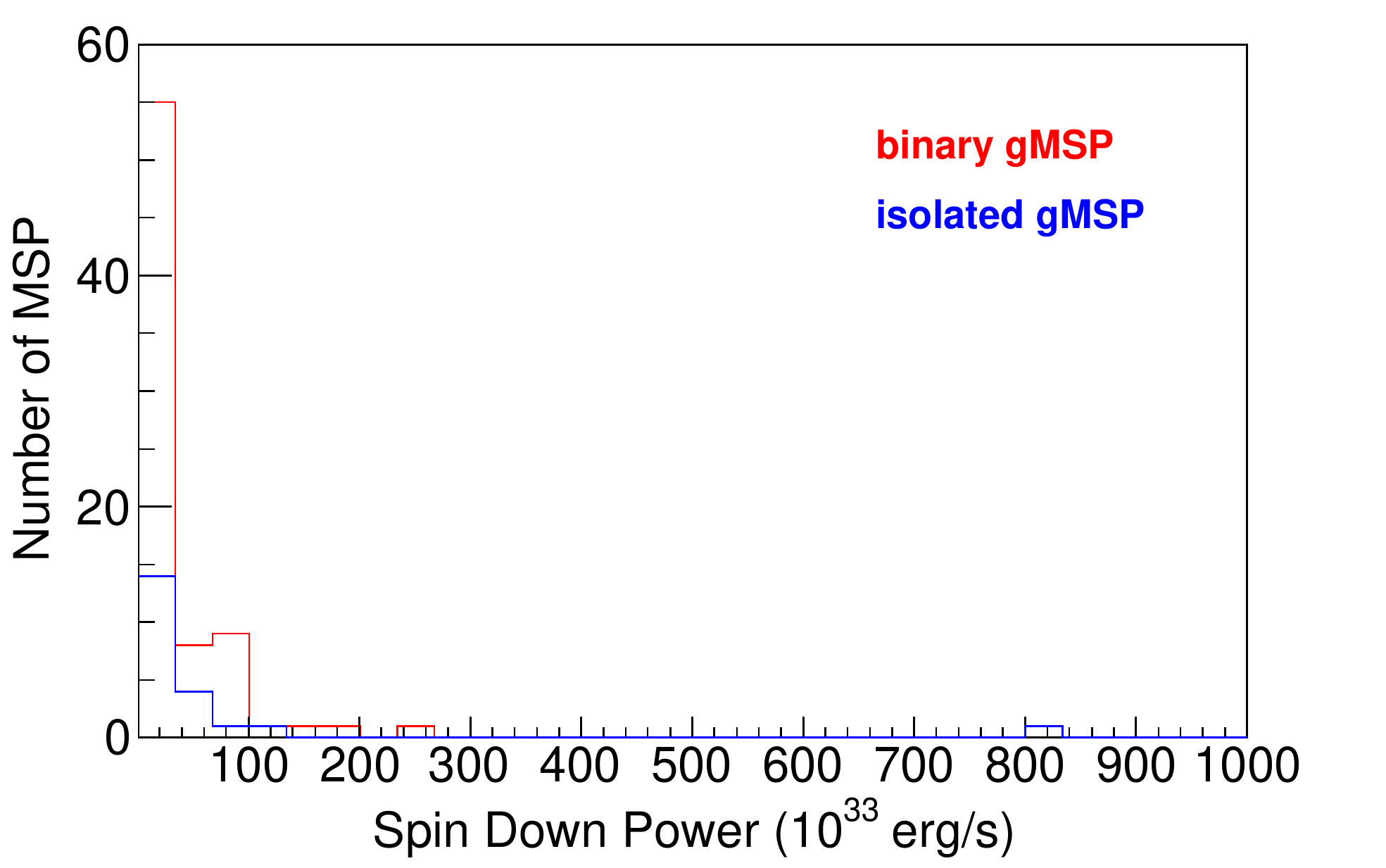}\\
\includegraphics[scale=.25]{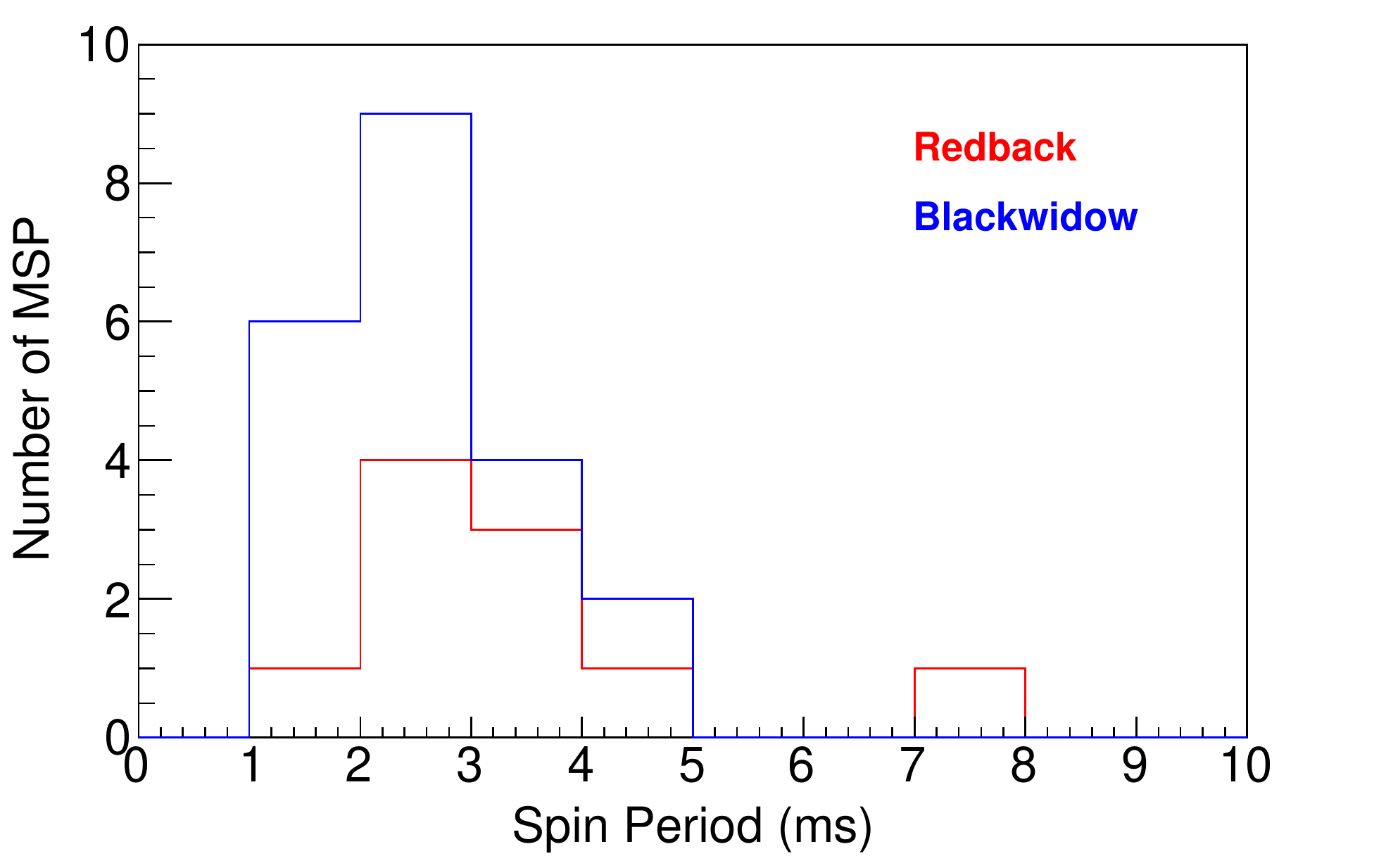}
\includegraphics[scale=.25]{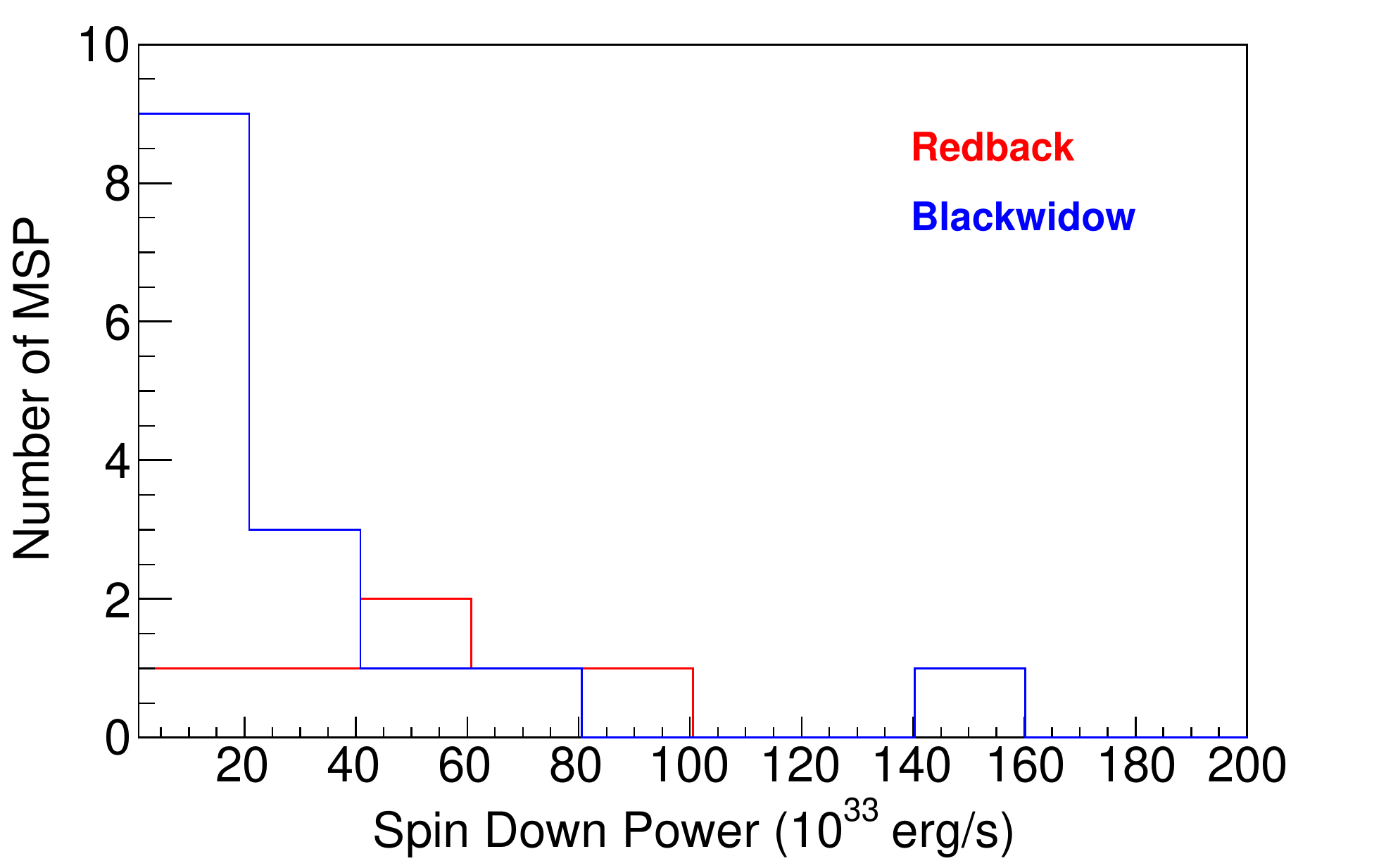}
\caption{Top:
Distribution of pulsar spin periods and spin down powers of the gMSPs with and without radio pulsation.
Middle: {\it ibid.} but for isolated gMSPs and gMSPs in binaries.
Bottom: {\it ibid.} but for redbacks and black widows.}
\label{map_radio}       % Give a unique label
\end{figure}

We note that most of the currently known 115 \fermi-LAT gMSPs are also detected having radio pulsations (see Table \ref{sum}).
We show the distribution of $P$  and $\dot{E}$ of gMSPs with and without radio pulsations in Figure \ref{map_radio}.
No difference can be drawn.
$\sim$80\% of \fermi-LAT detected gMSPs are in binary systems (Table \ref{sum}).
The distribution of $P$  and $\dot{E}$ comparing isolated gMSPs and gMSPs in binaries is also 
shown in Figure \ref{map_radio}.
The spin-down power distribution is consistent between these two populations.
However, the spin period of isolated gMSPs shows a hint for a bimodal distribution, which is different from that found of gMSPs in binary systems.
However, we caveat that the current population of isolated gMSPs is comparably small (24, Table \ref{sum}).
It is expected that a future catalog of \fermi-LAT gamma-ray pulsars may shed light on this point, having higher statistics.

Among MSPs in binaries, there are two interesting sub-groups known as redbacks (RBs) and black widows (BWs), on which we provide further details below.
They are tight binaries ($P_{orb} <$ 1 day) with a low mass companion star (BWs, $M_{companion}\ll0.1M_{Sun}$, e.g., \cite{Fruchter1988}; RBs, $M_{companion} \sim 0.2-0.4 M_{Sun}$, see e.g.,  \cite{Amico2001}).
In these cases, the strong pulsar wind is continuously ablating their companion star, leading to mass loss as indicated by the irregular eclipses of the radio pulsed
emission, which are in turn caused by the absorption and scattering of the ejected matter. 
Many of these systems have been discovered by radio surveys in our Galaxy (the ATNF Pulsar Catalog\footnote{\url{http://www.atnf.csiro.au/people/pulsar/psrcat/}}; \cite{Manchester2005}).
\fermi-LAT detected 10 RB and 21 BW (the Millisecond Pulsar Catalogue\footnote{\url{https://apatruno.wordpress.com/about/millisecond-pulsar-catalogue/}}).
The spin period distributions of these RBs and BWs are consistent, while the RBs may have a wider spin down power distribution than BWs (Figure \ref{map_radio}).
However, again, a large uncertainty still exists because of the small population considered, and these apparent differences have to be taken with extreme caution for the moment.

\begin{figure}[tb]
\centering
\includegraphics[scale=.25]{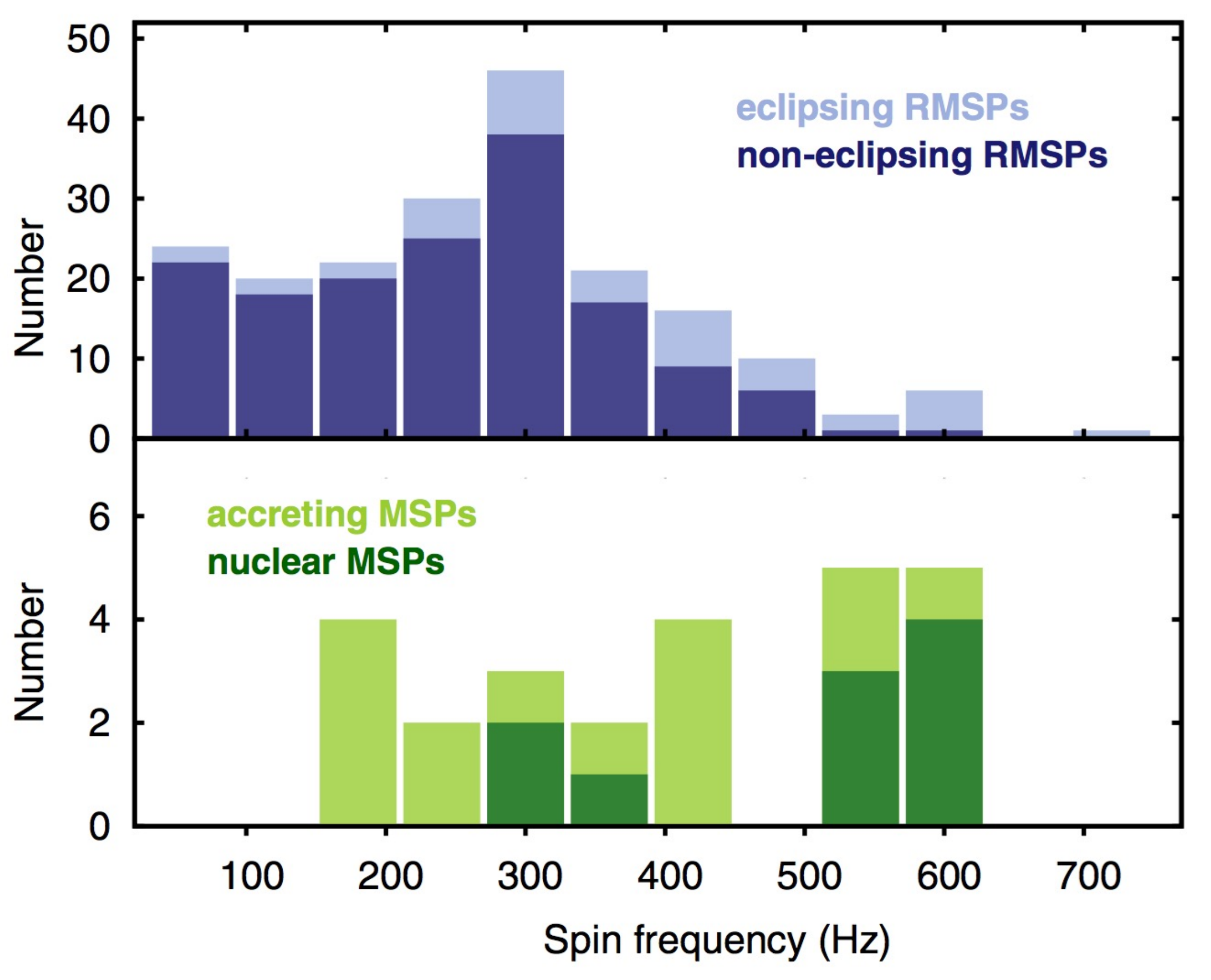}
\caption{Spin frequency distribution of binary non-eclipsing and eclipsing rotation-powered millisecond pulsars, and nuclear-powered and accreting millisecond pulsars. Taken from \cite{Papitto2014}.}
\label{spin}       % Give a unique label
\end{figure}

In a greater context, the comparison of the the spin frequency distributions of different samples of MSPs can be used to test theories 
 describing the spin evolution of NSs as they evolve from the accretion to the rotation-powered stages of the recycling scenario.
For instance, to study whether 
accreting MSPs are faster than binary millisecond radio pulsars (RMSPs), as noted by several authors (e.g., see \cite{Ferrario2007,Tauris2012} and references therein).
Considering the total population, and not just the gMSPs, the spin frequency distributions of binary millisecond pulsars
 have been recently studied by \cite{Papitto2014}.
They found that 
even though accreting MSPs are on-average faster than RMSPs by $\sim$100 Hz, comparing the two distributions gives a probability of $\sim0.4$
%0.374 
that they come from the same parent distribution. 
Simulations revealed that 
to detect a 3$\sigma$ difference to the rotational-powered MSP distribution there need to be at least 50 accreting MSPs. 
Assuming a normal distribution, 
this implies that if the spin frequency distributions of accreting MSPs and RMSPs are different, the difference would be visible when the number of known accreting MSPs has increased by more than a factor of three.
Only nuclear MSPs, i.e., nuclear-powered millisecond pulsars, with quasi-coherent-oscillations observed exclusively during thermonuclear type I X-ray bursts, are significantly faster than the rotational-powered MSPs.
Taking all of this into account, an obvious question arises.

%%%%%%%%%%%%%%%%%%%%%%%%%%%%%%%%%%%%%%%
\section{Do accreting millisecond pulsars shine in gamma-rays?}
%%%%%%%%%%%%%%%%%%%%%%%%%%%%%%%%%%%%%%%

Accretion-powered millisecond pulsars (AMSPs) normally orbit a low-mass companion star ($\sim$1 M$_\odot$ or less) and show coherent X-ray pulsations.
The latter are caused by the impact of an accretion stream onto the neutron star surface.
Thus, the coherent pulsations observed in the X-ray light curves (especially during outburst) occurs as a result of a kinetic, thermal process in nature, where 
at least part of the matter in the accretion flow is channeled to the surface of the star.
The prototypical source for this class is SAX J1808.4-3658, the first AMSP discovered \citep{Wijnands1998}.
Can such an accreting environment be also prone to the production of higher-energy gamma-rays?
How?
Can a radio and gamma-ray pulsar arise in periods of quiescence when accretion is not that dramatic?
These questions have unclear answers for the moment.

Despite radio or optical pulsations have not been detected from AMSPs up to the moment of writing,
see e.g., \cite{Burgay2003,Iacolina2010}
[the case of transitional pulsars is discussed in the next section],
some indications of an active pulsar state occurring in generic AMSPs have been put forward.
The amount of optical light reprocessed by the companion
star during X-ray quiescence in SAX J1808.4-3658
 \citep{Homer2001} is compatible
with irradiation by a radio pulsar \citep{Burderi2009}.
Also, 
the
decrease of the spin period of the neutron stars between consecutive outbursts is
similar to the rate observed from MSPs
\citep{Patruno2012}.
However, these are only indirect hints that such a scenario is not ruled out, not proof of its existence.

In gamma-rays, specially interesting due to their unscathed scape from significant absorption, 
the sky region of several AMSPs was investigated by \cite{Xing2013}, leading to no detections. 
The orbital uncertainties, as well as the 
limited time span of their search (4 years) could perhaps have played a role in their lack of detection of any candidate.
Indeed, the impact of the orbital uncertainties on the timing of pulsars in binary systems is large. 
\cite{Caliandro2012b} recently
presented an analytical study aimed to understand this impact on pulsation searches  for uncertainties in each of the orbital parameters 
and validated it with numerical simulations. 
Their table 3 summarizes the results.

\begin{figure}
\centering
\includegraphics[width=\linewidth, angle=0]{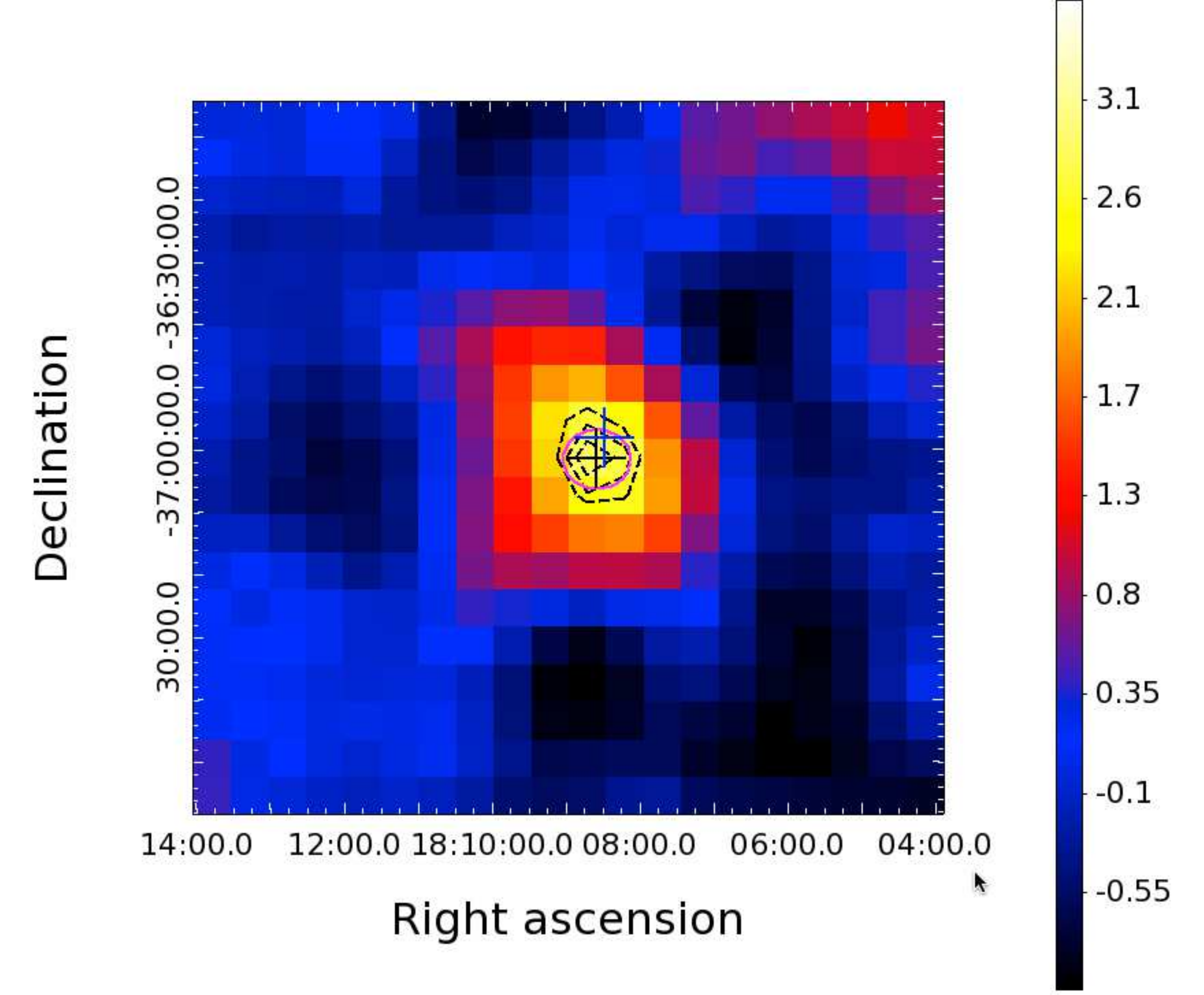}
\caption{ From \protect\cite{Wilhelmi2016}, using 6 years of data from the Large Area Telescope on board the Fermi Gamma-ray Space Telescope.
{\it Fermi}-LAT residual 2 deg$\times$2 deg (using a pixel size of 0.1 deg $\times$ 0.1 deg) count map above 1 GeV of the
  SAX J1808.4--3658 region smoothed with a Gaussian of width
  $\sigma=0.3$ deg (units of the scale on the right are
  counts). The best-fit position of the gamma-ray source is marked
  with a black cross whereas the position of SAX J1808.4--3658 is
  marked in blue. 
  }
\label{fig1}
\end{figure}

\cite{Xing2015b} later reported a barely significant modulation of SAX J1808.4-3658
at the spin-period, although 
perhaps the most relevant indication of gamma-ray emission from accreting millisecond pulsars is to be found in the work by
\cite{Wilhelmi2016}.
They revealed a point gamma-ray source at a significance of $\sim$6 $\sigma$ with a position compatible with that of SAX J1808.4-3658 within 95$\%$ confidence level (see Figure \ref{fig1}). 
Its flux and spectral parameters resulted compatible with that of 3FGL J1808.4-3703, a source in the 3FGL catalogue that is also compatible with the position of 
SAX J1808.4-3658.
\cite{Wilhelmi2016} could not claim an association by timing, though, given that 
the search for gamma-ray pulsations is limited by the uncertainties in the ephemeris and position of the source as well as by its dim character.
However, they did rule out the existence of  other possible obvious candidates to producing the gamma-rays observed beside the AMSP. 
If proven true, SAX J1808.4--3658 in X-ray quiescence will be similar to, for instance, PSR J1311-3430 \citep{Pletsch2012,Ray2013}, also a 
fast MSP ($\sim$2.5 ms) in a compact binary system ($\sim$2 h). The spectral parameters are also compatible within the current statistics to the ones found in other MSPs, with a hard
spectrum and a turn over at a few GeV.

If the detection of SAX J1808.4-3658 is confirmed, the implication is that the AMSP would be active in gamma-rays along the whole quiescent state.
The fully opposite alternative from associating gamma-ray emission to ASMPs could come from the possibility that the former is transient, and perhaps linked with 
outburst events.

A recent search for gamma-ray emission from AMSPs from this perspective have been presented by \cite{Xing2019}.
They reported hints for detection of GRO J1008-57, a Be X-ray binary with 
transient X-ray pulsar, detectable during X-ray outbursts of the source, having a period of 93.5 s \citep{Coe2007}.
Interestingly, the magnetic field of the pulsar is known to be the highest among the Be XRBs, likely as high as $8 \times 10^{12}$ G \citep{Shrader1999}.
GRO J1008-57 shows type-I X-ray outbursts at each periastron passage (e.g., \cite{Tsygankov2017}), produced by the 
interaction between the neutron star and the circumstellar disk around the Be companion (e.g., \cite{Reig2011}), 
and also, occasionally, type-II outbursts. 
\cite{Xing2019} proposed that the source also occasionally flares in gamma-rays, following giant X-ray outbursts. 
\cite{Li2012} have also earlier proposed a similar case for the transient gamma-ray behavior 
detected as GRO J1036-55 and AGL J1037-5708 to be associated with the HMXB 4U 1036-56.
In fact, their table 1 shows a list of unidentified transient gamma-ray sources in the Galactic plane with possible binary nature.
However, although not yet ruled out, evidence for an association is circumstantial in all these cases, since at the end of the day, it is mostly based on positional coincidence:
the occurence of the gamma-ray emitting events did not precisely coincide with any X-ray outbursts involved.
In the case of GRO J1008-57, not even a uniform displacement (all after, or all before) of the putative gamma-ray emission and the outburst is found (see, e.g.,
figure 5 of \cite{Xing2019}).

From a theoretical perspective, it is unclear how an accreting scenario could lead to gamma-ray pulsations.
Based on \cite{Cheng1991}, \cite{Romero2001} suggested that hadronic process, 
 in which hadrons accelerated from the magnetosphere of the neutron star could impact the surrounding accretion disk, could lead to observable gamma-ray emission. They applied this model to 
A 0535+26 back in EGRET times. 
However, no further {\it Fermi}-LAT detection of the source has been found, and 
additionally, in this model one would also expect a significant neutrino yield \citep{Anchordoqui2003}
which has not been detected either, see e.g., \citep{Acciari2011,Aartsen2017}.
\cite{Bednarek2009,Bednarek2009b} proposed that gamma-rays can be produced at a turbulent and magnetized transition region formed due to the balance between the magnetic pressure and the pressure injected by the accreting matter. 
Depending on the accretion level, matter can either accrete or be expelled away, similarly to a propeller. 
However, it is unclear in these models how the pulsations could be maintained, and in particular, how could a magnetosphere sustain penetration of accreting matter without being disrupted.
We discuss more on this in the next section, related to transitional pulsars.

The recent radio detection of
an evolving jet from a strongly-magnetised accreting X-ray pulsar have been also put forward
\citep{vandenEijnden2018,Russell2018}.
Whereas the nature of origin of the radio emission (whether in a jet, or otherwise) may be yet unclear, the fact that there is likely non-thermal emission in these environments 
is not.
The same region of non-thermal particle acceleration can also be prone to gamma-ray production.

At this time, however, we can safely conclude that it is yet unclear whether gamma-ray emission from purely AMSPs has or has not been detected yet. 
The dim character of the expected emission and slow increase in signal-to-noise ratio with time in a gamma-ray background-dominated domain play against a ready confirmation. 
Further observations of the most promising candidates are needed in order to reach to a definite claim.
In particular, ASMP transient emission would be certainly detectable (provided it reaches to sufficiently high energy, of course) by the Cherenkov Telescope Array \citep{2019scta.book}.
The capabilities of the latter in comparison with {\it Fermi}-LAT are much better for shorter events, e.g., it can be up to tens of thousands of times for sensitive for events lasting an hour or less.
This capability may open a yet-unchartered transient gamma-ray sky.

%%%%%%%%%%%%%%%%%%%%%%%%%%%%%%%%%%%%%%%
\section{Gamma-ray emission from transitional millisecond pulsars}
%%%%%%%%%%%%%%%%%%%%%%%%%%%%%%%%%%%%%%%

%
PSR J1023+0038 \citep{Archibald2009,Stappers2014} was the first source for which two different states (accretion and rotational) were suggested.
Later on, swings between a rotation-powered MSP state and an accretion-powered low mass X-ray binaries (LMXB) state were caught  
on a few-weeks timescale in the transient system IGR J18245-2452 \citep{Papitto2013}.
Such timescales could then be compatible with those of the accretion flow.
XSS J12270-4859 \citep{deMartino2010,deMartino2013,Bassa2014,deMartino2015}, has also been observed to transition.
These systems are called transitional pulsars.
The phenomenology of these transitions is varied and complex, and there is a whole chapter in this book dealing with that, which should be read concurrently with this section to
get a full panorama.
We shall  only provide here some comments describing the multi-wavelength phenomenology, focusing on what relates to describing the gamma-ray properties of these sources.

We have clearly defined states 
at the two extremes of the transition:
On the one hand, at high mass inflow rates, the radio pulsar is  likely 
shut-off, the system is bright in X-rays ($L_X > 10^{36}$ erg s$^{-1}$). 
This is an accretion state, similar to other AMSPs in LMXB.
On the other hand, at sufficiently 
low mass inflow rates, the magnetosphere likely dominates the scenario and is in full fledge up to the light cylinder, the radio pulsar is active, 
the disk disappears and
the system is instead dimmer in X-rays ($L_X \sim 10^{32}$ erg s$^{-1}$).

\begin{figure}[t]
\centering
\includegraphics[width=0.96\textwidth,angle=-90]{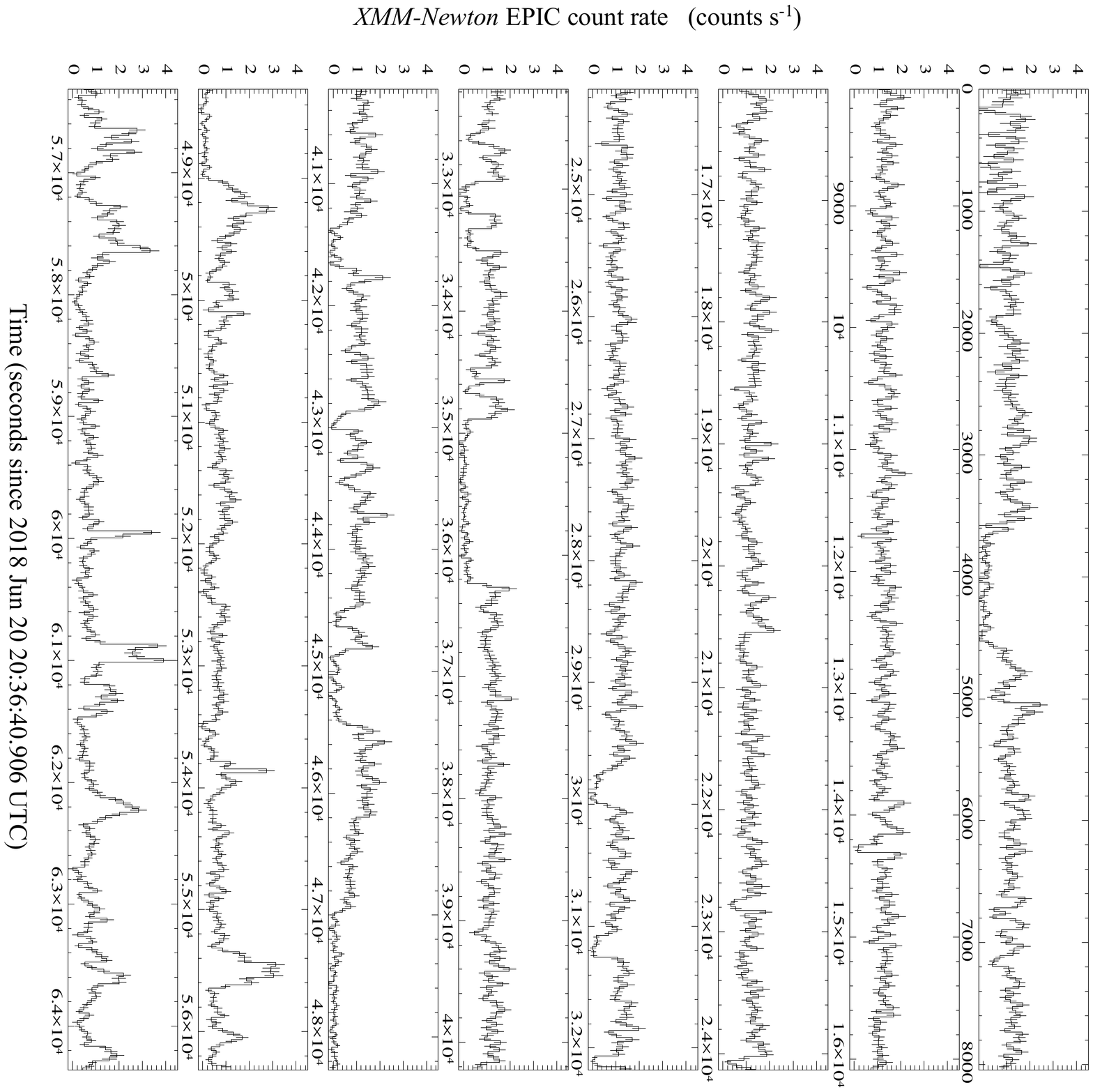}
\caption{ An example of a sub-luminous state for a transitional pulsar.
0.3-10 keV background-subtracted and exposure-corrected light curve of CXOU J110926.4-650224, as obtained with XMM–Newton EPIC with a time bin of 30 s.
Talen from \protect\cite{CotiZelati2019}. }
\label{sublu}
\end{figure}

The most interesting behavior happens in between these two extremes, where all three sources mentioned have been observed to be in a sub-luminous accretion state, 
with X-ray luminosities in between (of the order of ($L_X \sim 10^{33}$ erg s$^{-1}$).
In this sub-luminous state (also called, active state)  
these systems vary their X-ray fluxes from a low to a high mode (spending most of the time in the latter) with occasional flaring. 
Whereas the states (both the radio pulsar and sub-luminous state) can be stable along years, 
the mode-changing within the sub-luminous state, which has been seen in all transitional sources, 
can be very fast, and has been often found on a time scale of a few seconds. 
What causes such rapid changes is not yet well understood.
Figure \ref{sublu} provides a graphical example of the evolution of the X-ray luminosity in one of the transitional pulsars candidates, CXOU J110926.4-650224.
General features regarding the variability are generic in all members of the class.
In fact, such features of the sub-luminous state are the defining characters of transitional pulsars as we understand them now, although the community is still calling them candidates
until the rotational state is observed (what may well never happen in our lifetime, given the decades-long duration of the sub-luminous states in the prototypical sources).
At least four candidates to transitional pulsar systems have been identified.
\cite{CotiZelati2019} list their properties (see their Table 6, as well as provide detailed references for them all).
They all share in the general features of the sub-luminous states, with their high and low modes, but have not been observed to undergo a state transition so far.

\begin{figure}[t]
\centering
\includegraphics[width=0.492\textwidth]{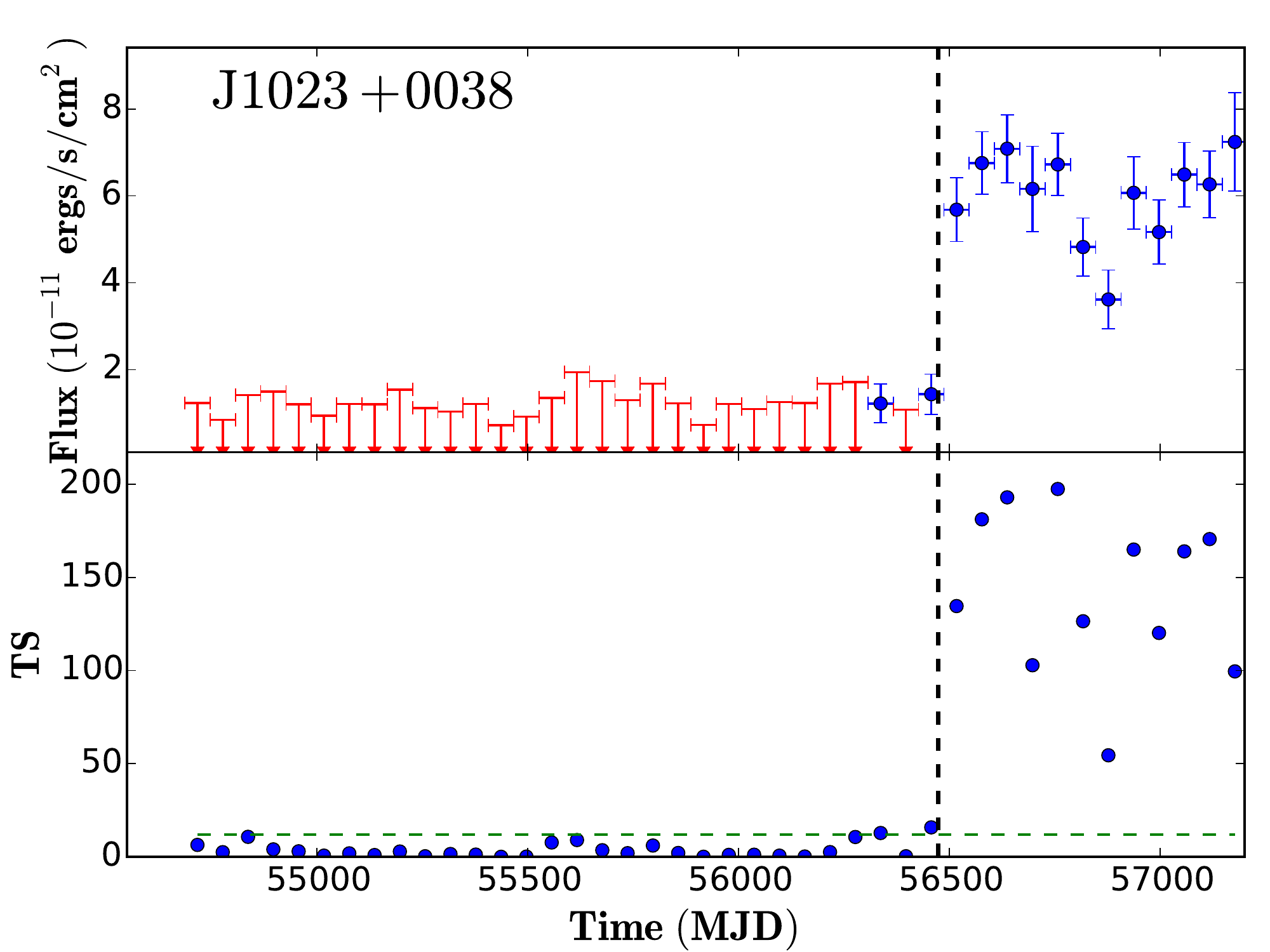}
\includegraphics[width=0.492\textwidth]{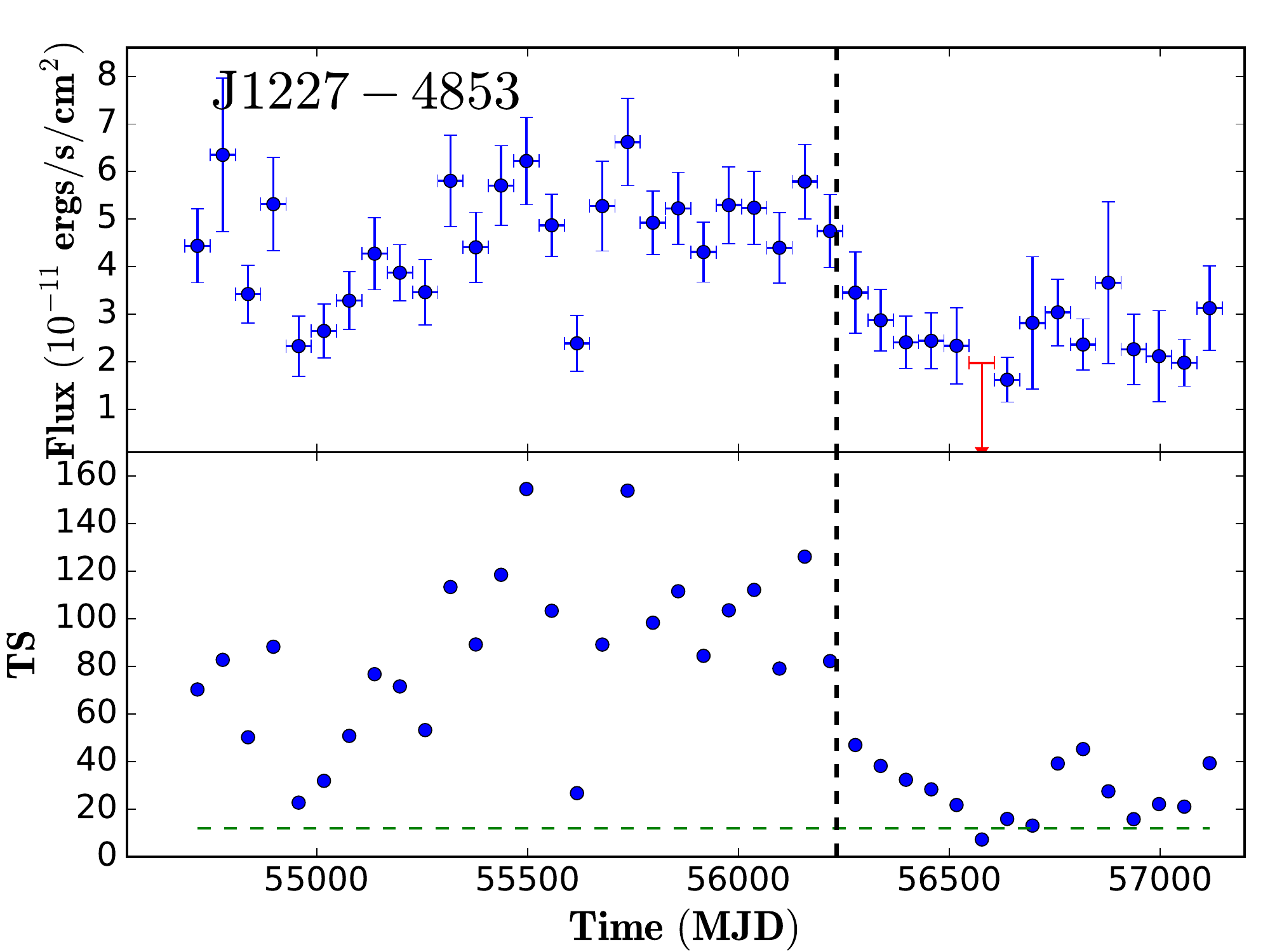}
\includegraphics[width=0.492\textwidth]{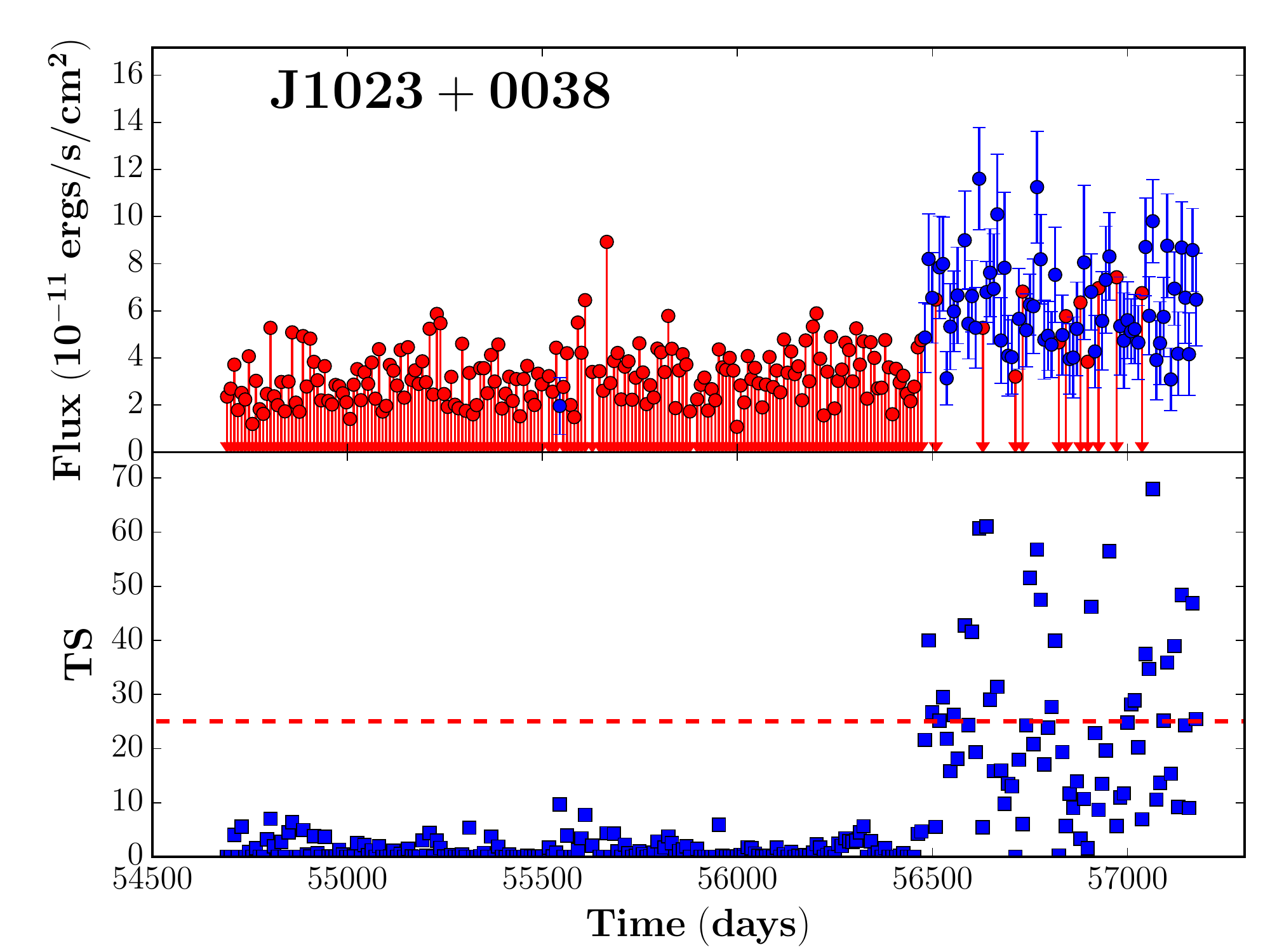}
\includegraphics[width=0.492\textwidth]{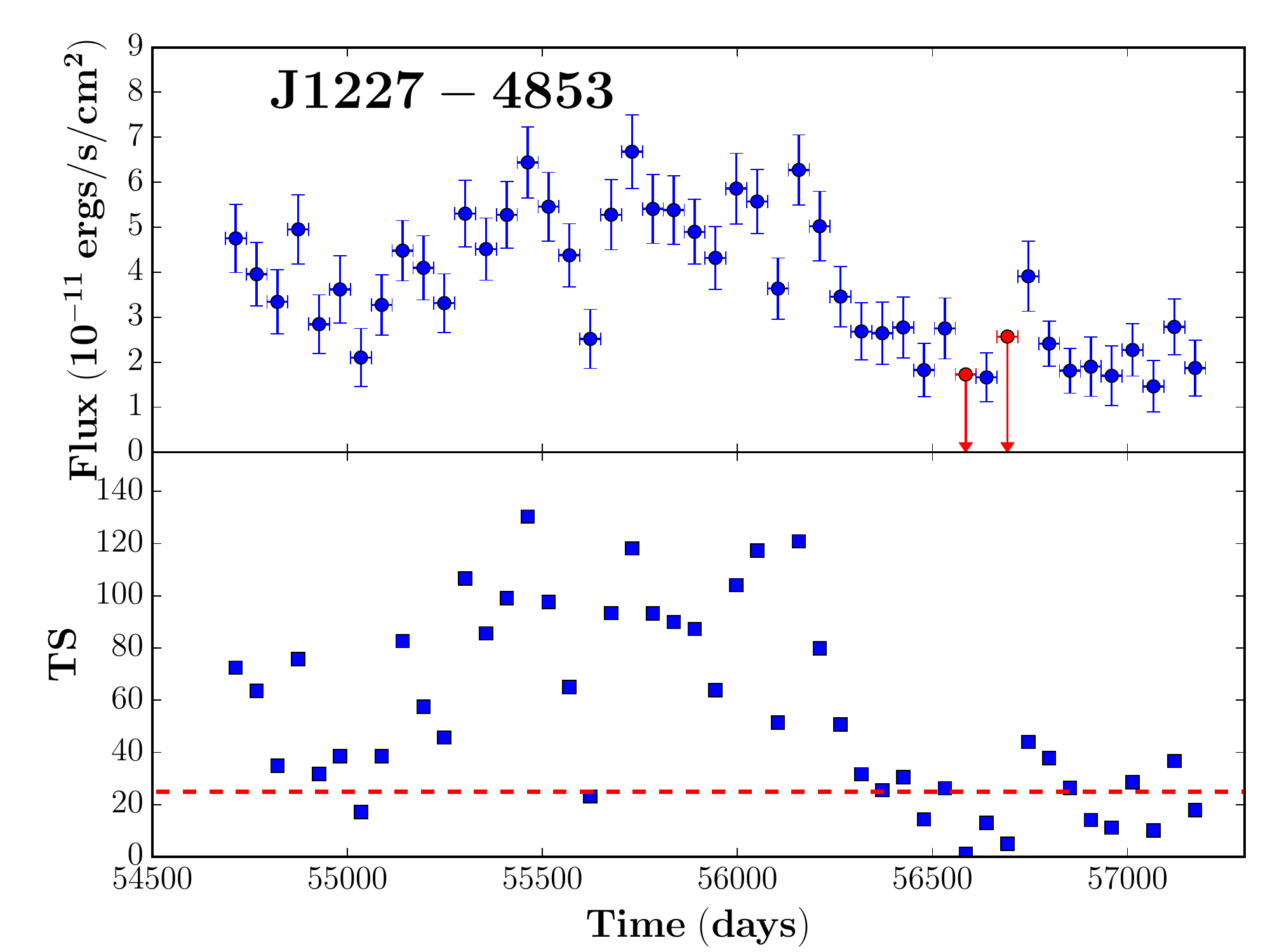}
\includegraphics[width=0.492\textwidth]{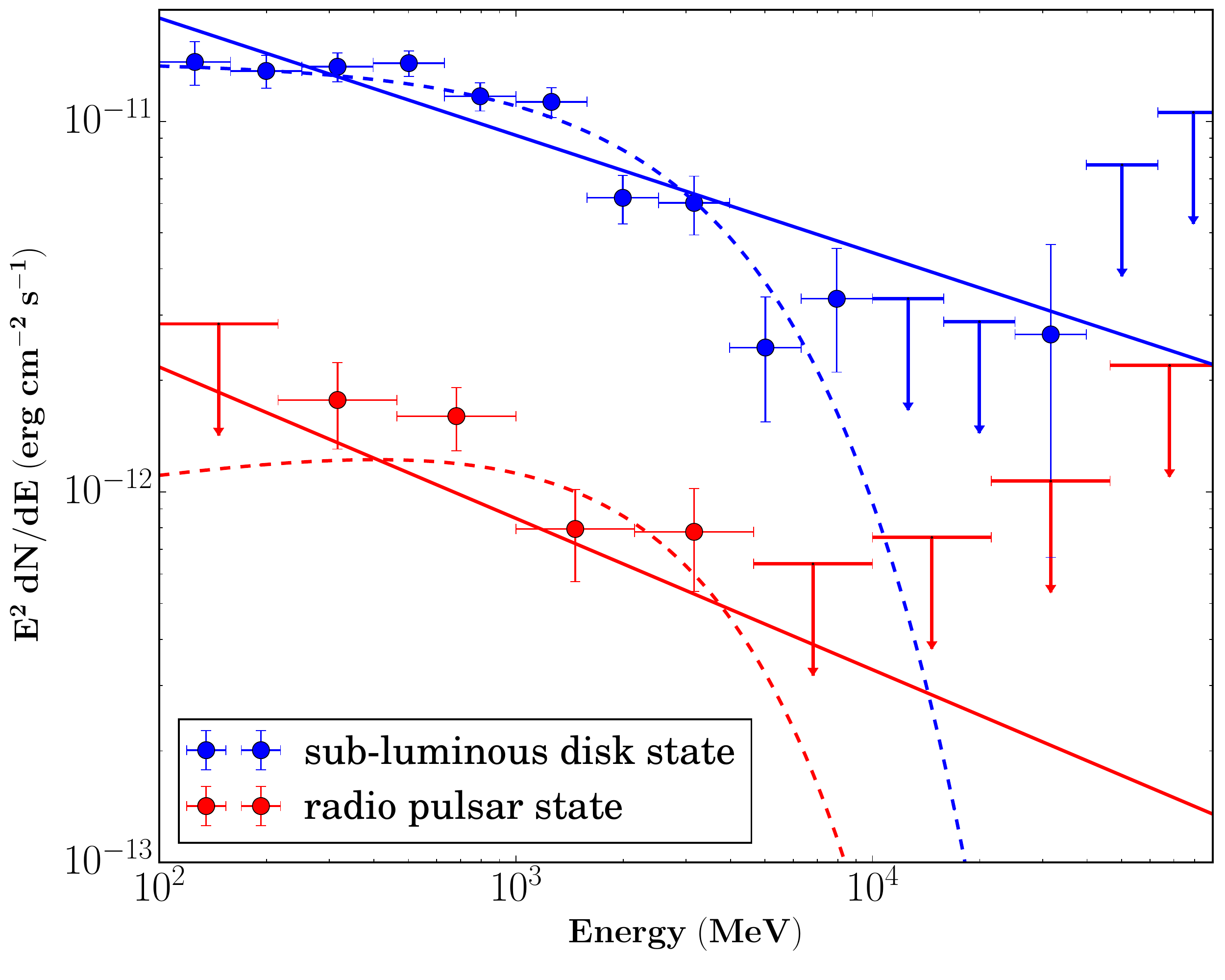}
\includegraphics[width=0.492\textwidth]{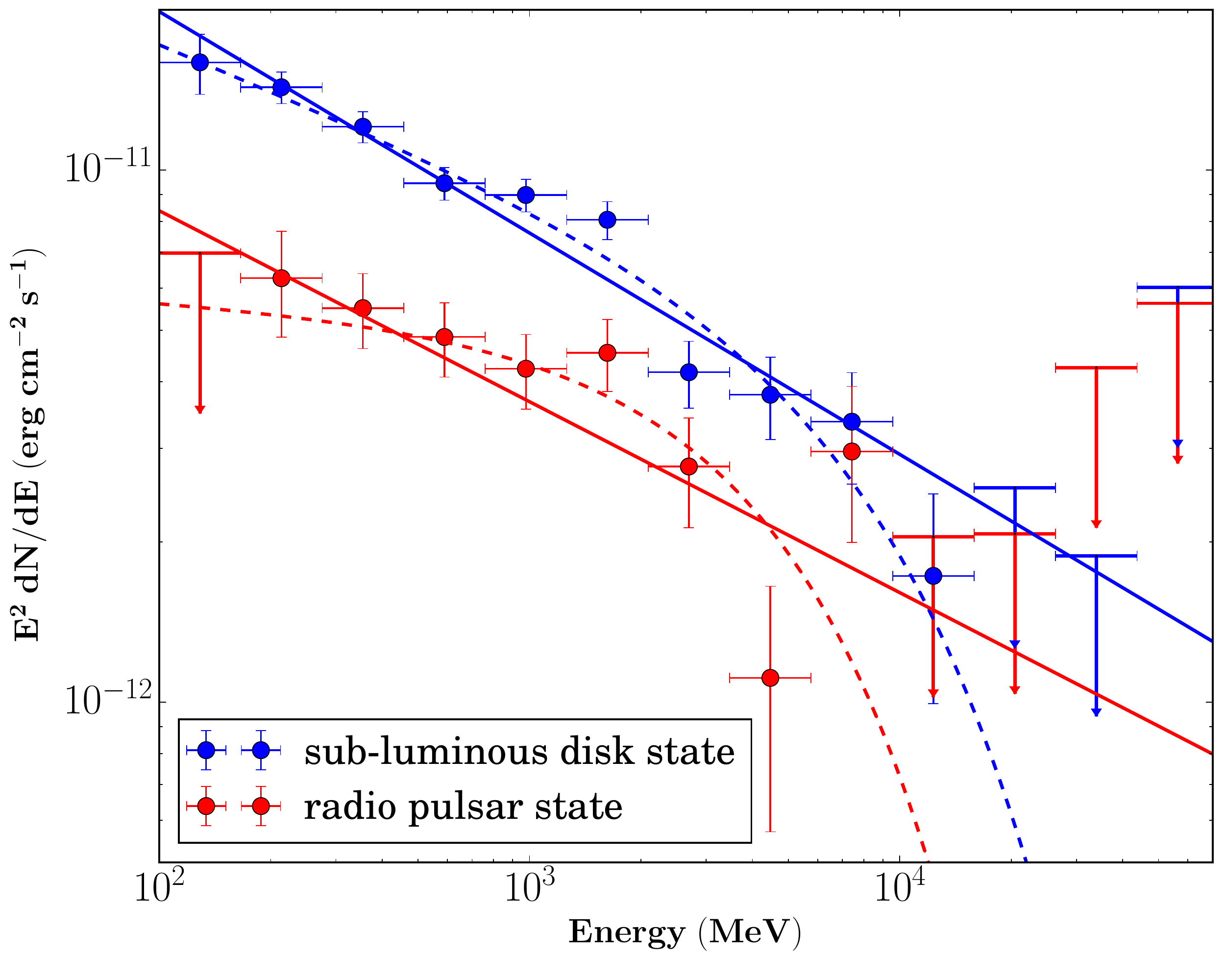}
\caption{Adapted 
from \protect\cite{Torres2017}. The first row shows the long-term light curves of the transitional millisecond pulsars J1023+0038 and J1227$-$4853. The already known (Stappers et al. 2014; Bassa et al. 2014) state transitions are indicated with dotted vertical lines. The red lines show the flux upper limits. The dotted horizontal green line indicates TS=12. 
The second row shows the same data but analyzed using a smaller time bin defined by simulations.
The third row shows the \emph{Fermi}-LAT spectra of J1023+0038 (left) and J1227$-$4853 (right), in radio pulsar state (red) and sub-luminous disk state (blue).
}
\label{LC-2}
\end{figure}

It is interesting to note that the gamma-ray emission also transitions from quiescence to the active state simultaneously with the X-rays and optical flux.
For instance, in the cases of J1023+0038 
\citep{Stappers2014}; and J1227-4853 \citep{Johnson2015} (by the way, both being RB systems), 
their gamma-ray flux was observed to vary by a factor of 2 to 5.
Recently, \cite{Torres2017} considered 7 years of \fermi-LAT data to search transitions between states in all
redbacks and black widows.
Figure \ref{LC-2} shows the long-term light curves and the spectra of the transitional millisecond pulsars J1023+0038 and J1227$-$4853. 
Regarding the spectra, high-energy cutoffs at a few GeV are stablished for the high gamma-ray state of PSR J1023+0038 and for both states of PSR J1227-4853.
In the light curves, the state transitions are indicated with dotted vertical lines. The red lines in Figure \ref{LC-2}
show the flux upper limits. The dotted horizontal green line indicates TS=12 (which for \fermi-LAT is roughly equivalent to about 3.5 $\sigma$ confidence level).
In the analysis by \cite{Torres2017}, the smallest time bin so that a state transition similar to those found in the known transitional pulsars would be detected with high confidence was defined for each
system analysed. 
For instance, in the case of J1023+0038 and J1227$-$4853 this time bin is 8 and 53 days, respectively,
and the light curve under these bins are shown in the second row of Figure \ref{LC-2}.
With this method, it can be seen that whereas 
in the low state, the flux evolution is compatible with being constant,
the high gamma-ray state conflicts with a constant flux.
For instance, for J1227$-$4853, it is ruled out with a significance of 4.7$\sigma$, likely indicating the action of shorter-timescale phenomena also in gamma-rays.
Indeed, under the current understanding of transitional pulsars, it is expected that in the low mode of the sub-luminous state the rotationally-powered MSP is active again (and with it, the gamma-ray emission).
If this is true, by adjoining all low-mode periods of the sub-luminous state one should be able to acquire a gamma-ray signal. This signal should in principle be the pure contribution of the MSP.
This test is however hampered by the fact that one would need a constant monitoring to determine all the low-mode periods in the light curve, given that with the historic simltaneous X-ray/gamma-ray data, adjoining all 
such already-determined periods does not allow for enough statistics to be acquired. No significant upper limit can be put.

Using the same binning method referred above, \cite{Torres2017} found no hint for a state transition for most of the studied pulsars, 
two black-widow systems, PSR J2234+0944 and PSR J1446-4701 have an apparent variability that is reminiscent of the transitions in PSR J1023+0038 and PSR J1227-4853.

%
%

%%%%%%%%%
% Pulsations in transitional systems? [modify text]
%%%%%%%%%

Evidence for gamma-ray pulsations in J1023+0038 during the quiescent state has been presented by 
\cite{Archibald2013}.
They presented evidence 
at the 3.7$\sigma$ confidence level
(using gamma-ray data from 2008 to 2012, before the transition).
Currently in the active state, pulsations remain hidden.
Additionally, 
X-ray pulsations at the neutron star spin period have been earlier observed to exist both in the quiescent state 
%at a confidence level of 4.5$\sigma$ \citep{Archibald2010},
%
and in the high mode of the active state 
\citep{Archibald2015}.
These high-mode pulsations were initially understood in terms of intermittent, and partial, mass accretion onto the NS surface.
The 1.69 ms spin period of PSR J1227-4853 (in system XSS J12270-4859) was also recently discovered in gamma-rays, once the source transitioned to 
a rotation-powered MSP, as inferred from decreases in optical, X-ray, and gamma-ray flux from the source  \citep{Johnson2015}. 

Transitional pulsars are then the only low-mass X-ray binaries from which emission at energies of few GeVs has been undoubtedly observed.
The immediate interpretation that has been put forward is
that the gamma-ray emission is due to a disk/pulsar wind shock, similar to gamma-ray binaries (see below),
see e.g., \cite{Stappers2014,Li2014b,Takata2014,CotiZelati2014}.
In these scenarios, a rotation-powered pulsar must be active even in the presence of an
accretion disk, when the radio coherent pulsation is washed out by the
enshrouding of the system by intra-binary material. 
In more detail, 
\citet{Stappers2014} and \citet{CotiZelati2014} proposed
that the pulsar wind/mass in-flow shock is the region where the gamma-ray emission
is generated.
Instead, \citet{Takata2014} and
\citet{Li2014b} interpreted the gamma-ray emission through inverse Compton scattering of UV disk
photons directly by the wind.
Here, the
X-rays are due to synchrotron emission taking place in the shock
between the pulsar wind and the plasma in-flow. 
Such a shock would be
expected to be stronger in the sub-luminous disk state when a larger fraction of the pulsar wind would be
intercepted.

However, the observation of coherent X-ray pulsations --with an rms amplitude of $\approx 6$ per
cent during the accretion disk state-- is difficult to reconcile with these interpretations. 
If these are rotation-powered, then when a disk is present  the pulsed flux increases more than one order of
magnitude with respect to the case of an unperturbed magnetosphere.
What makes this happen?
In fact, one would expect that the larger the plasma density, the 
more shorted out the electric fields which power the electron/positron acceleration
should be.
Also the total energy reservoir would be in doubt.
The sum of the average luminosity observed from J1023+0038 in just the 0.3--79 keV
and 0.1--100 GeV energy bands amounts to $\simeq 1.7\times10^{34}$ erg
s$^{-1}$, a value that implies a spin-down power conversion efficiency
of $>40\%$. 
Depending on how the source emits in the 1-10 MeV energy range, where we
lack observational tools, the  total
luminosity would  
significantly exceed the spin down power.
Strong flickering (as observed in X-rays) makes the case
for the spin-down power being the lone source of energy even more
unlikely.

Models in which most of the matter in the disc would have to be propelled away 
were also constructed \citep{Ferrigno2014,Papitto2014,Papitto2015,Campana2016}, and have shown to be 
successful in describing the gamma-ray emission. 
In here, the
emission is due to self-synchrotron Compton
emission originated at the turbulent boundary between a propelling
magnetosphere and the disk in-flow.
At lower energies, the
  X-ray emission is contributed by synchrotron emission
   from the same region, and by the luminosity emitted by
  the accretion flow. 
  The  inability of observationally  separating  these contributions  limits the model testing.
However, the recent detection of optical pulses (during high and flare modes) at a flux level compatible with a power law
extrapolation from X-rays 
\citep{Ambrosino2017,Papitto2019} is intriguing.
The pulsed luminosity observed in the visible band is too large to be produced by reprocessing of 
accretion-powered X-ray emission or cyclotron emission by electrons in the accretion columns above the pulsar polar caps.
Instead, it was interpreted 
as an indication that a rotation-powered pulsar was active.
It is not yet settled 
whether optical pulses are produced in a magnetospheric environment (as in other millisecond pulsars) or are perhaps the result of 
a mini pulsar wind nebula. 
In the latter scenario, proposed by Papitto et al. (2019), the striped pulsar wind 
meets the accretion disk within a few light cylinder away from the pulsar, and this intrabinary shock provides the synchrotron radiation.

%%%%%%%%%%%%%%%%%%%%%%%%%%%%%%%%%%%%%%%
\section{More on redbacks, and black widows in the context of gamma-ray binaries}
%%%%%%%%%%%%%%%%%%%%%%%%%%%%%%%%%%%%%%%

Gamma-ray binaries are binary systems producing most of their electromagnetic output in gamma rays above 1 MeV (see \citep{Dubus2015} for a review).
Their multi-wavelength emission are orbitally modulated from radio to TeV.
Additionally, in at least one case, a super-orbital modulated signal has been found \citep{Li2012c,Chernyakova2012,Li2014c,Ackermann2013,Ahnen2016b,Chernyakova2017}.
There are only a handful {of} known
gamma-ray binaries: six in the Galaxy: PSR B1259-63 \citep{Abdo2011, Aharonian2005a, Caliandro2015}; 
\lsi\/, \cite{Abdo2009a,Albert2006,Hadasch2012}; 
LS 5039, \cite{Abdo2009b,Aharonian2005b,Aharonian2006,Collmar2014,Hadasch2012}; 
1FGL J1018.6-5856, \cite{Abramowski2015,Ackermann2012a,Li2011a}; 
\hess\/, \cite{Aharonian2007, Aliu2014,Bongiorno2011,Li2017}; 
4FGL J1405.1-6119, \cite{Corbet2019});
and one in the Large Magellanic Cloud (CXOU J053600.0-673507, \cite{Corbet2016}).
The currently known gamma-ray binaries are all high mass X-ray binary systems, hosting a massive O or Be star and a compact object.
Except for PSR B1259-63, hosting a 48~ms pulsar, the nature of the compact objects in such binaries is unknown.
Colliding wind (pulsar / stellar wind) interaction (e.g., \cite{Maraschi1981,Dubus2006}), pulsar wind zone processes (e.g., \cite{Bednarek2011,Bednarek2013,Sierpowska2008},
a transitioning pulsar scenario (e.g., \cite{Zamanov2001,Torres2012,Papitto2012}), and microquasar jets (see e.g., \cite{Bosch2009} for a review) have  been  proposed  as  the  origin of the gamma-ray emission for one or several gamma-ray binaries.
None of these gamma-ray binaries have a known pulsar with a period classifying it as MSP. 
There might be chance, however, that some of the binaries with a yet unknown companion has a millisecond pulsar in the system, although it would have to be a newborn 
neutron star, given the short-lived nature of the companion. 
Other X-ray binary systems also show gamma-ray emission.
Transient gamma-ray emission from V404 Cyg has been observed by \fermi-LAT in coincidence with the brightest radio and hard X-ray flare during its 2015 June outburst \citep{Loh2016}.
Cyg X-1 \citep{Albert2007,McConnell2000,Sabatini2010} and Cyg X-3 \citep{Abdo2009c,Corbel2012,Tavani2009} have also been detected in gamma rays.
However, their spectral energy distributions (SEDs) peak at {X-ray energies}, and their gamma-ray emission is not recurrent in every orbit.
Thus, we do not consider V404 Cyg, Cyg X-1 and Cyg X-3 as gamma-ray binaries (and of course, none seem to contain a millisecond pulsar as compact object).

\begin{figure}[tb]
\centering
\includegraphics[scale=.28]{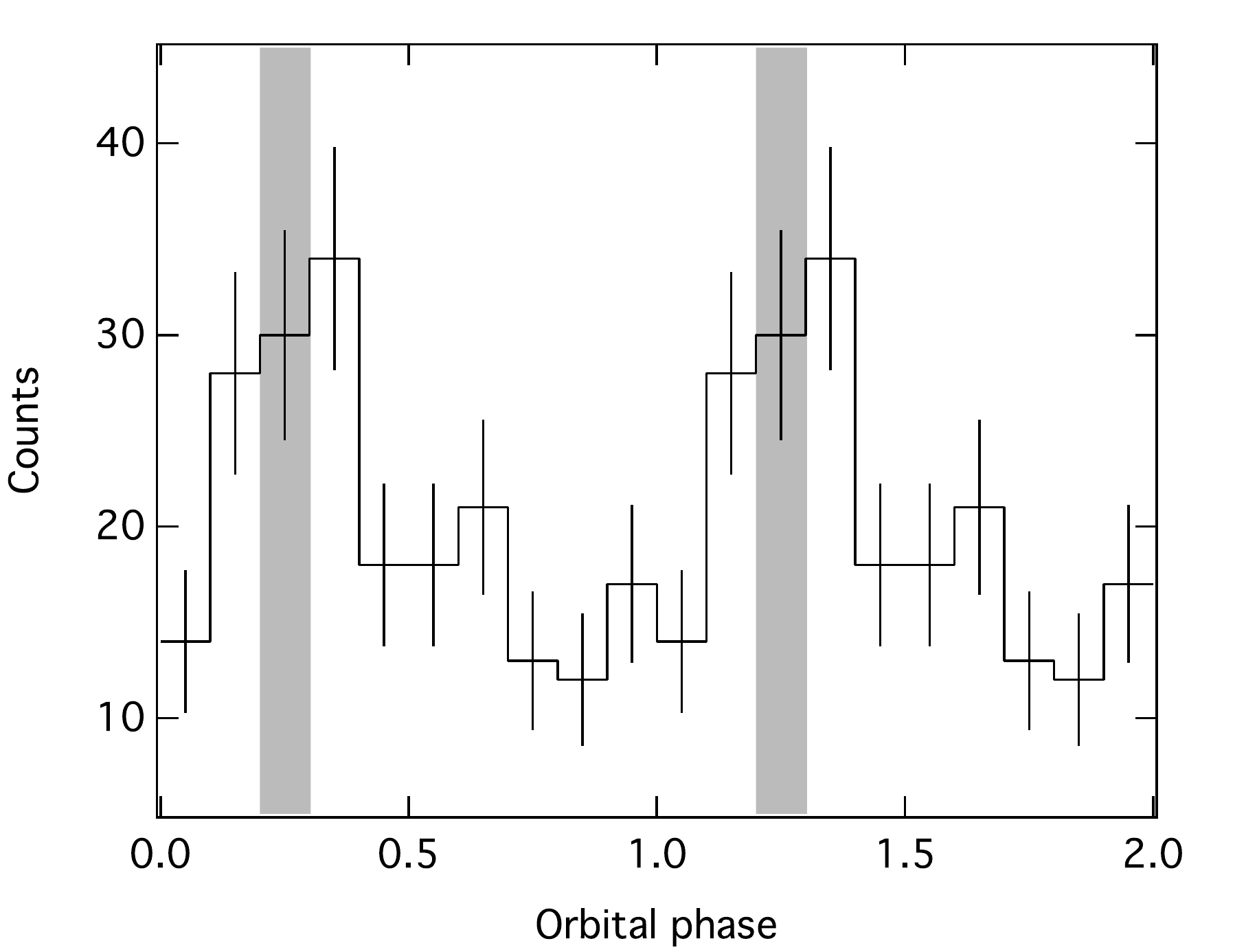}\\
\includegraphics[scale=.32]{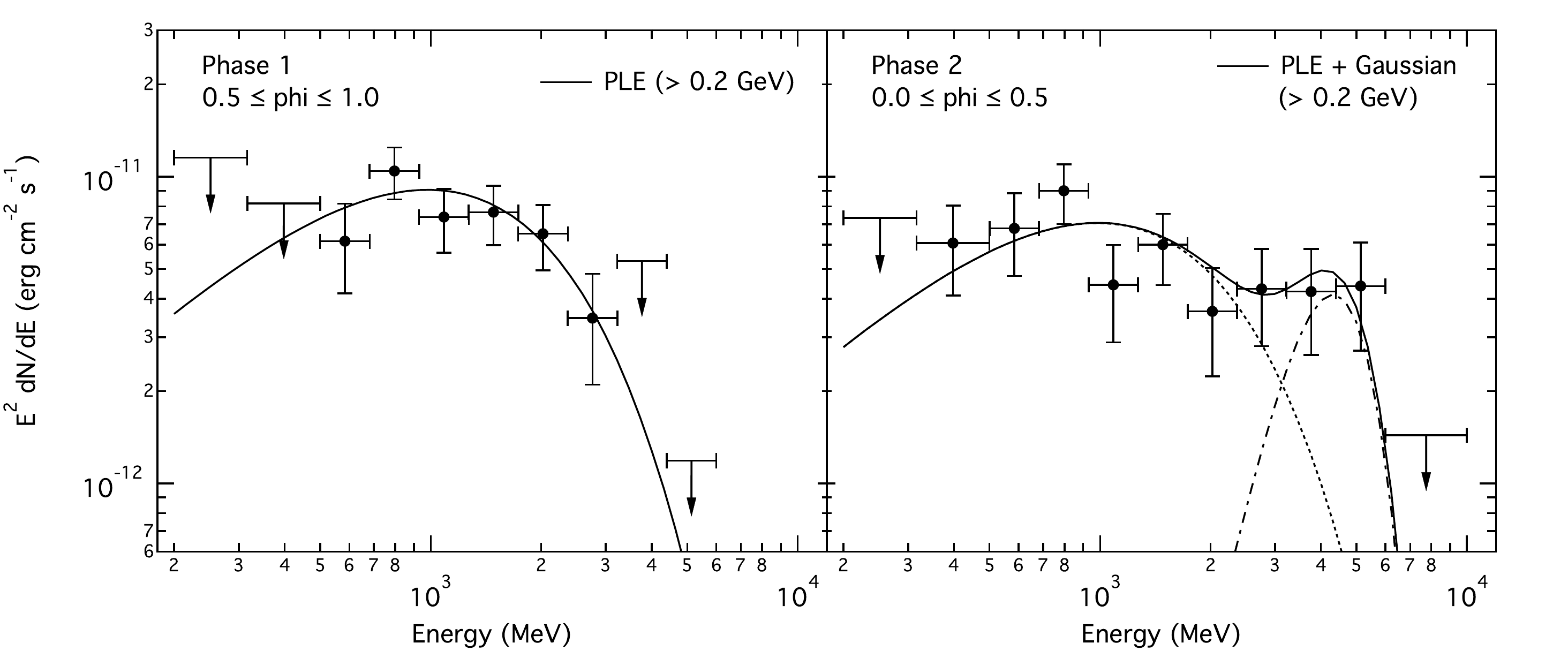}
\caption{Top: gamma-ray light curve of PSR B1957+20 above 2.7 GeV folded at the orbital period. Bottom: gamma-ray spectrum for the low (orbital phase 1, 0.5-1.0) and high (orbital phase 2, 0.0-0.5) states. 
Taken from \cite{Wu2012}.}
\label{WU}       % Give a unique label
\end{figure}

In RBs and BWs, 
the collision between stellar wind and pulsar wind also produces and intra-binary shock, which is thought to accelerate particles to relativistic energies similarly as in gamma-ray binaries.
Synchrotron emissions from these intra-binary shocks are observed in X-rays having orbital modulation, both for RBs and BWs (e.g., see \cite{Gentile2014, Roberts2015, Roberts2018}), while gamma-rays are expected to arise through inverse Compton scattering (e.g., see \citep{Tavani2009, Bednarek2014}).
X-ray emission of about half of the BWs are dominated by the thermal emission from MSP surface and did not reveal much shock emission.
Some BWs show a clear non-thermal X-ray component but except in some cases, e.g. PSR B1957+20 \citep{Huang2012}, 
their orbital light curves are not clearly defined due low statistics.
RBs, on the contrary, nearly always show a clear non-thermal X-ray spectra components and orbital modulated light curve, also thought to be driven by the intra-binary  shock (see e.g., \cite{Roberts2018}).
In RBs and BWs, the intra-binary shock induced gamma-ray emission manifests as orbital modulation in light curves and spectra.
For instance, the gamma-ray orbital light curve of BW PSR B1957+20 above 2.7 GeV from \cite{Wu2012} is shown in Figure \ref{WU}.
The modulation is apparent to eye.
The corresponding gamma-ray spectra for the low (orbital phase 1, 0.5-1.0) and high (orbital phase 2, 0.0-0.5) states are shown in Figure \ref{WU}.
An orbital phase related spectral modulation is detected (\cite{Wu2012}).

The search for intra-binary shock-induced gamma-ray emission in RBs and BWs is difficult since gMSP themselves are much brighter gamma-ray sources.
To minimize the contamination of gMSPs, a search during their off-peak phases is preferred.
Valid timing ephemerides is then a prerequisite for the search,  but the timing noise of gMSPs, orbital solution uncertainties and orbital period fluctuation of RBs \& BWs make the task difficult.
Out of 10 RBs and 21 BWs detected by \fermi-LAT, there are only 2 RBs and 4 BWs showing hints (above 3$\sigma$) of intra-binary shock-induced gamma-ray emission (Table \ref{RB_BW} for details and references).
Besides that, 2 RBs candidates also show hints of orbital modulated gamma-ray emission.
In RBs and BWs, the currently observed X-ray and gamma-ray emissions from intra-binary shock resemble those in gamma-ray binaries, peaking above 1 MeV.
In this perspective, RBs and BWs in Table \ref{RB_BW} are potential gamma-ray binaries.
However, the current gamma-ray orbital modulation of RBs and BWs are all below 5 $\sigma$.
Future observations are needed to increase the significance and confirm their classification as gamma-ray binaries.

In fact, we note that a number of still unidentified \fermi-LAT sources may host a BW or a RB pulsar, 
which could have been not yet detected as such because it was not observed in the radio band, 
or because its radio signal was scattered and absorbed by matter enshrouding the system, 
or because no X-ray variable counterpart has been discovered yet.

\begin{table*}{}
\centering
\caption{RBs and BWs with intra-binary shock-induced gamma-ray emission}
\begin{tabular}{cccl}
\\
\hline\hline
\\
  Name    &   RB/BW   &  Orbital modulation            &  References \\
                 &                  &    in light curve                            &    \\

    \\
\hline\hline
\\
PSR B1957+20    &  BW    &  $\sim 2.3\sigma$  & \cite{Wu2012}   \\
\\
PSR J1311-3430 & BW    & $\sim 3.5\sigma$   & \cite{Xing2015}; \cite{An2017}  \\
\\
PSR J0610-2100 & BW  & $\sim 2\sigma$        &  \cite{Espinoza2013MNRAS} \\
\\
PSR J2241-5236 & BW  & $\sim 4.4\sigma$     &  \cite{An2018} \\
\\
PSR J1023-0038 & RB &    $\sim 3.2\sigma$      &   \cite{Xing2018} \\
\\
PSR J1227-4853 &  RB   &  $\sim 3.0\sigma$   & \cite{Xing2015} \\
\\
2FGL J0523.3-2530 &  RB candidate &  $\sim 4.0\sigma$   & \cite{Xing2014} \\
\\
 3FGL J2039.6-5618 &  RB candidate &  $\sim 4.0\sigma$  & \cite{Ng2018} \\
\\
\hline\hline\
\\
\\

%\tablecomments {The first (second) uncertainties correspond to statistical (systematic) errors.}
\label{RB_BW}
\end{tabular}
\end{table*}

%%%%%%%%%%%%%%%%%%%%%%%%%%%%%%%%%%%%%%%%%%%%%%%%%%
\subsection*{Closing comments: MSPs at home}
%%%%%%%%%%%%%%%%%%%%%%%%%%%%%%%%%%%%%%%%%%%%%%%%%%

As closing comments we would like to draw attention to the radio-quiet MSPs.
As we have seen, essentially 
all rotationally powered MSPs have been discovered through their radio pulsations.
This is of course 
limiting our knowledge of the population to those MSPs that are nearby, bright and 
with radio emission beamed toward us.
Whereas it is true that 
MSPs are expected to have wider radio beams, visible from a larger range of viewing angles, 
thus 
making radio-bright pulsars the most common 
(see, e.g., the population studies by \cite{Story2007}), there is also an obvious observational bias.
The problem is that
precise position-dependent barycentering corrections need to be applied to photon arrival times to account for the Doppler shift due to \fermi’s motion through the solar system. 
Without the radio detection, the localization of a gamma-ray source suspected to be a pulsar (e.g., see \cite{SazParkinson2016}) cannot be better than a few arcmin, much larger than the arc second precision required to detect gamma-ray pulsations from MSPs. Hundreds of thousands of sky locations covering the source localization region must therefore be searched.
This tremendous computational effort has been outsourced 
on the distributed volunteer computing system Einstein@Home \citep{Knispel2010}.
This allowed 
to search for pulsations from more than 150 pulsar-like unidentified sources (those having curved spectra and low flux variability).
This survey discovered several MSPs, now even a radio quiet MSPs, see \cite{Clark2018}, promoting the hope of obtaining
a more unbiased sample.

%%%%%%%%%%%%%%%%%%%%%%%%%%%%%%%%%%%%%%%%%%%%%%%%%%
\subsection*{Acknowledgments}
%%%%%%%%%%%%%%%%%%%%%%%%%%%%%%%%%%%%%%%%%%%%%%%%%%

We acknowledge support from grants PGC2018-095512-B-I00, SGR2017-1383, and AYA2017-92402-EXP. 
Jian Li further acknowledges the support from the
Alexander von Humboldt Foundation. 
We also acknowledge the fruitful discussions with the international team on 
“Understanding and unifying the gamma rays emitting scenarios in high mass and low mass X-ray binaries” of the ISSI (International Space Science Institute), Beijing.

%%%%%%%%%%%%%%%%%%%%%%%%%%%%%%%%%%%%%%%%%%%%%%%%%%

\end{document}